\documentclass[12pt]{article}

\usepackage[hmargin=2.7cm, vmargin=2.7cm]{geometry}
\usepackage{amsmath,amssymb,amsthm}
\usepackage{actuarialsymbol}
\usepackage{subcaption}
\usepackage{multirow}
\usepackage{bm}
\usepackage{graphicx}
\usepackage{caption} 
\usepackage{subcaption} 
\usepackage{enumitem} 
\usepackage{booktabs}
\usepackage{colortbl} 
\usepackage{array}
\usepackage{float}
\usepackage{hyperref} 
\usepackage[dvipsnames]{xcolor}
\hypersetup{ 
    colorlinks = true,
    linkcolor = black, 
    urlcolor=blue,
    citecolor = black,
    pdfpagemode=FullScreen}
\usepackage{soul}
\usepackage{natbib}

\usepackage{setspace} 
\newcommand{\mcP}{\mathcal{P}}
\newcommand{\mcA}{\mathcal{A}}

\newcommand{\mcI}{\mathcal{I}}
\newcommand{\mcL}{\mathcal{L}}

\newcommand{\bbE}{\mathbb{E}}
\newcommand{\VaR}{\text{VaR}}

\makeatletter

\definecolor{LightBlue}{RGB}{140,186,252}
\definecolor{LightYellow}{RGB}{253,253,150}
\definecolor{newred}{RGB}{156, 34, 54}

\definecolor{newblue}{RGB}{93, 63, 211}

\begin{document}

\begin{center}

\textbf{\large A Natural Hedging Framework for Longevity Risk with Graphical Risk Assessment}

~\\

Lydia J. Gabric$^{a}$ and Kenneth Q. Zhou$^{b,}$\footnote{Corresponding author. E-mail: {\tt kenneth.zhou@uwaterloo.ca}}
\end{center}

\begin{center}
$^{a}$School of Mathematical and Statistical Sciences, Arizona State University, USA \\
$^{b}$Department of Statistics and Actuarial Science, University of Waterloo, Canada \\
\end{center}

\noindent\textbf{Abstract:} Natural hedging allows life insurers to manage longevity risk internally by offsetting the opposite exposures of life insurance and annuity liabilities. Although many studies have proposed natural hedging strategies under different settings, calibration methods, and mortality models, a unified framework for constructing and evaluating such hedges remains undeveloped. While graphical risk assessment has been explored for index-based longevity hedges, no comparable metric exists for natural hedging. This paper proposes a structured natural hedging framework paired with a graphical risk metric for hedge evaluation. The framework integrates valuation, calibration, and evaluation, while the graphical metric provides intuitive insights into residual dependencies and hedge performance. Applied to multiple hedging scenarios, the proposed methods demonstrate flexibility, interpretability, and practical value for longevity risk management.

\vspace{0.3cm}\noindent\textit{Keywords:} Natural hedging; Longevity risk; Stochastic mortality models; Graphical risk metric; Hedge effectiveness

\section{Introduction} \label{sec:introduction}

    The uncertainty surrounding future mortality trends, known as longevity risk, has been extensively studied in the context of life insurance valuation and risk management. In general, there are two categories of longevity risk management solutions: internal and external. External solutions transfer longevity risk from life insurers to capital market investors through index-based hedging instruments and standardised contracts. Internal solutions, in contrast, utilize the insurer's own portfolio structure to mitigate longevity and mortality risks. Among these, natural hedging has received considerable attention for its intuitive mechanism of balancing opposite exposures within life insurance and annuity portfolios.

    The seminal work of \citet{Cox2007} introduced the concept of natural hedging by leveraging the inverse relationship between life insurance and annuity products to diversify mortality and longevity risks. When mortality changes, the values of life insurance and annuity liabilities typically move in opposite directions, allowing the insurer to offset these changes and reduce overall risk exposure. Since \citet{Cox2007}, numerous studies have explored how this relationship can be more effectively utilized and implemented in different settings to manage mortality and longevity risks.
    
    For instance, \citet{Lin2014} proposed an immunization-based approach to construct natural hedges using sensitivity measures. \citet{Zhu2014} explored a non-parametric mortality model and found that natural hedging performance may be less effective compared with parametric models. More recently, \citet{yang2019} combined internal natural hedges with external index-based hedges under a unified framework. \citet{Chen2024} analysed product demand and supply considerations in natural hedging, while \citet{cupido2024} studied the impact of spatial dependence on hedge performance. These studies collectively demonstrate the growing importance of natural hedging in longevity risk management.

    Despite the aforementioned advancements, several challenges remain in constructing natural hedging effectively. First, identifying appropriate life insurance products to offset the mortality risk of annuity products can be complex. Second, determining the optimal quantity of life insurance products to effectively hedge requires detailed mathematical calibration. Third, evaluating hedge effectiveness is complicated due to residual risks such as calibration risk, structural basis risk, and model risk. Although prior research has attempted to address aspects of these challenges, there is still no unified framework that allows the systematic construction, calibration, and comparison of natural hedges across different models and methods.

    In addition, existing studies primarily rely on numerical risk measures to evaluate hedging performance, which may not fully capture the joint behaviour of liabilities and hedging instruments. This shortfall could understate the strength of the inverse relationship or the offsetting effects in natural hedging. To address these gaps, this paper develops a unified natural hedging framework that integrates both numerical and graphical evaluation methods. The framework provides a consistent process for valuation, calibration and evaluation, while the accompanying graphical risk metric offers visual insights into the resulting hedge performance. The contributions of this paper are threefold.

    The first contribution of this paper is the development of a risk management framework specifically designed for natural hedging. While frameworks for constructing and evaluating external longevity hedges have been proposed in prior studies, such as the index-based hedging procedure of \citet{coughlan2011longevity} and the decomposition of hedge effectiveness in \citet{longevity_hedge_effective_decomp}, a comparable framework dedicated to natural hedging has not been established. To fill this gap, we propose a three-step hedging framework that extends the structure of \citet{longevity_hedge_effective_decomp} and adapts it for natural hedging.

    The proposed framework consists of three stages: valuation, calibration, and evaluation. In the first stage, we derive formulas for valuing life insurance and annuity products under a stochastic mortality setting. In the second stage, we consider multiple hedge calibration techniques categorized by their objectives, including variance minimisation, duration matching, and delta hedging. In the third stage, hedge evaluation combines risk measures with graphical tools to provide a comprehensive and interpretable assessment of hedging effectiveness. The integration of graphical evaluation within the framework leads to our second contribution.

    The second contribution of this paper is the introduction of a graphical risk metric specifically tailored to natural hedging. Graphical tools have been explored in the context of standardised longevity hedging, such as the graphical basis risk metrics proposed by \citet{graphical_basis_risk_measure} and extended by \citet{grm_applied_sherris}, as well as in \citet{Blake2008}, \citet{Dowd2010} and \citet{Li2021}. Since no comparable metric has been developed for evaluating natural hedges, we propose a graphical risk metric that visualizes the relationship between life annuity and insurance portfolios, allowing for a direct and intuitive evaluation of hedge performance.
    
    The proposed graphical risk metric is constructed using a series of joint prediction regions that represent potential hedge outcomes across multiple confidence levels. To facilitate practical application, we further develop an interpretation procedure that enables users to visually assess hedge effectiveness and compare alternative hedging strategies. Visual elements such as shaded regions and reference lines highlight deviations from expected outcomes and the degree of diversification effects. This graphical approach provides a transparent and intuitive means to diagnose whether hedging shortfalls arise from under-hedging, over-hedging, or a lack of hedging effect.

    The third contribution of this paper is the numerical implementation of the proposed natural hedging framework and graphical risk metric across three practical applications. First, we demonstrate how the framework can identify trade-offs when selecting an insurance portfolio as the hedging instrument. Second, we compare alternative calibration techniques to determine the optimal allocation of the hedging portfolio. Third, we evaluate model risk by comparing two stochastic mortality models and a non-parametric approach for a given portfolio configuration. The results from these illustrations show that the proposed framework and graphical risk metric together serve as a flexible and interpretable tool for constructing and evaluating natural hedges.

    The remainder of the paper is organized as follows. Section~\ref{sec:framework} outlines the components of the natural hedging framework and demonstrates its use through a toy example. Section~\ref{sec:grm} describes the construction and interpretation of the graphical risk metric. Section~\ref{sec:illustrations} applies the framework and graphical metric to three practical illustrations. Finally, Section~\ref{sec:conclusion} concludes the paper and discusses possible extensions and limitations.

\section{The Natural Hedging Framework} \label{sec:framework}

    Consider a life insurer issuing both life annuity and life insurance products. Let $\mcA$ and $\mcI$ denote the present value random variables of the annuity and insurance portfolios, respectively. To hedge the longevity risk of $\mathcal{A}$, the insurer employs a natural hedging strategy resulting in a hedged position
    \[
    \mcP(h) = \mathcal{A} + \mathcal{L}(h), 
    \]
    where $\mathcal{L}(h) = h \cdot \mathcal{I}$ represent the calibrated life insurance portfolio and $h$ is the hedge ratio indicating the amount of $\mathcal{I}$ needed for the hedged position.

    To support a systematic implementation of such natural hedging, we develop a three-step framework in this section. We briefly summarise the three steps below:
    \begin{enumerate}
        \item \emph{Valuation of $\mathcal{A}$ and $\mathcal{I}$}: Derive the present value of the annuity and insurance liabilities, and obtain their expected values under stochastic mortality assumptions.
        \item \emph{Calibration of $h$}: Determine the hedge ratio that achieves a specified objective, such as variance minimization, duration matching, or delta neutral.
        \item \emph{Evaluation of $\mathcal{P}(h)$}: Assess the performance of the hedged portfolio through both numerical and graphical metrics, such as variance, Value-at-Risk, or histograms.
    \end{enumerate}
    The rest of this section examines each step of the proposed framework in detail and concludes with a simple illustrative example.

\subsection{Valuation} \label{sec:FrameworkValuation}    

    To construct a natural hedge, the insurer must first quantify the present value of the annuity and insurance liabilities. Under stochastic mortality modelling, these liabilities are random variables since their values depend on uncertain future survival outcomes. Thus, the first step of our natural hedging framework is to define and value these present value random variables as the foundation for establishing the hedge.

    Let $\mathcal{S}_{x}(T)$ be the probability that an individual aged $x$ at time $0$ will survive another $T$ years. This probability can be expressed as
    \begin{equation} \label{eq: gen_survival_prob}
        \mathcal{S}_{x}(T) = \prod_{s=1}^{T} (1 - q_{x+s-1, s}),
    \end{equation}
    where $q_{x, s}$ is the probability that an individual aged $x$ at the beginning of year $s$ dies during year $s$. In a stochastic setting, $\mathcal{S}_{x}(T)$ is a random variable since $q_{x,s}$, for $s=1,\ldots,T$, is uncertain due to random mortality fluctuations. We denote its expectation by $S_{x}(T) := \mathbb{E}[\mathcal{S}_{x}(T)]$. The procedure for generating realisations of $\mathcal{S}_{x}(T)$ is provided in Appendix~\ref{append:mort_generator}.
    
\subsubsection{Annuity liabilities}

    Recall that we denote $\mathcal{A}$ as the present value random variable of the annuity portfolio. If the insurer has $n_{A}$ distinct annuity products in the portfolio and applies a constant force of interest $\delta$, then $\mathcal{A}$ is given by
    \begin{equation*}
        \mathcal{A} = \sum_{j=1}^{n_{A}} \omega_{j}^{A} \sum_{k = \tau_{j}}^{\tau_{j} + t_{j} - 1} c_{j} \cdot e^{-\delta k} \cdot \mathcal{S}_{x_{j}}(k),
    \end{equation*}
    where $\omega_{j}^{A}$ is the weight of the $j$-th annuity product, $\tau_{j}$ is its deferral period, $t_{j}$ is the number of payments, $c_{j}$ is the payment size, and $x_{j}$ is the issuing age. The weights satisfy $\sum_{j=1}^{n_{A}} \omega_{j}^{A} = 1$ and $0 < \omega_{j}^{A} \leq 1$ for all $j$.

    It follows that the expectation of $\mathcal{A}$ represents the expected present value of the entire annuity portfolio liabilities, which can be written as
    \begin{equation} \label{eq: epv_annuity_portfolio}
        A := \mathbb{E} \left[ \mathcal{A} \right] = \sum_{j=1}^{n_{A}} \omega_{j}^{A} \sum_{k = \tau_{j}}^{\tau_{j} + t_{j} - 1} c_{j} \cdot e^{-\delta k} \cdot S_{x_{j}}(k).
    \end{equation}
    The distribution of $\mathcal{A}$ and the value of $A$ can be obtained using Monte Carlo simulations under stochastic mortality modelling.

\subsubsection{Insurance liabilities}

    Recall again that we denote $\mathcal{I}$ as the present value random variable of the insurance portfolio. If the insurer has $n_{I}$ distinct insurance products in the portfolio and applies a constant force of interest $\delta$, then $\mathcal{I}$ is given by
    \begin{equation*}
        \mathcal{I} = \sum_{j = 1}^{n_I} \omega_{j}^{I} \sum_{k=0}^{t_{j}-1} b_{j} \cdot e^{-\delta(k+1)} \cdot \left( \mathcal{S}_{x_{j}}(k) - \mathcal{S}_{x_{j}}(k+1) \right),
    \end{equation*}
    where $\omega_{j}^{I}$ is the weight of the $j$-th insurance product, $t_{j}$ is its term length, $b_{j}$ is the benefit amount, and $x_{j}$ is the issuing age. The weights again satisfy $\sum_{j=1}^{n_{I}} \omega_{j}^{I} = 1$ and $0 < \omega_{j}^{I} \leq 1$ for all $j$.
    
    Similar to the annuity liabilities, the expectation of $\mathcal{I}$ is the expected present value of the entire insurance portfolio liabilities, which can be written as
    \begin{equation} \label{eq: epv_insurance_portfolio}
        I := \mathbb{E}[\mathcal{I}] = \sum_{j = 1}^{n_I} \omega_{j}^{I} \sum_{k=0}^{t_{j}-1} b_{j} \cdot e^{-\delta(k+1)} \cdot \left( S_{x_{j}}(k) - S_{x_{j}}(k+1) \right).
    \end{equation}
    The distribution of $\mathcal{I}$ and the value of $I$ also can be obtained using Monte Carlo simulations under stochastic mortality modelling.

\subsubsection{Summary of notations}

    Having derived the annuity and insurance portfolios, we now summarise the notations before presenting the next two steps of the framework. The hedge ratio is $h$, the calibrated insurance portfolio is $\mcL(h) = h \cdot \mcI$ with its expected present value being $L(h) = h \cdot I$, and the combined portfolio (i.e., the hedged position) is $\mcP(h) = \mcA + \mcL(h)$ with its expected present value being $P(h) = A + L(h)$. For ease of reference, Table~\ref{tab:Valuation} provides a summary of the notations introduced for the natural hedging framework.
    \begin{table}[!ht]
        \centering
        \begin{tabular}{l | l l l }
            \toprule
             Description & Present Value (PV) & Expected PV & Mean-adjusted PV \\
             \midrule
             Annuity portfolio & $\mathcal{A}$ & $A$ & $\tilde{\mcA} = \mcA - A$\\
             Insurance portfolio & $\mathcal{I}$ & $I$ & $\tilde{\mcI} = \mcI - I$\\
             Calibrated portfolio & $\mathcal{L}(h)$ & $L(h)$ & $\tilde{\mcL}(h) = \mcL(h) - L(h)$ \\
             Combined portfolio & $\mcP(h)$ & $P(h)$ & $\tilde{\mcP}(h) = \mcP(h) - P(h)$ \\
             \bottomrule
        \end{tabular}
        \caption{Summary of the notations used in the natural hedging framework.}
        \label{tab:Valuation}
    \end{table}

\subsection{Calibration} \label{sec:FrameworkCalibration}

    We now turn to the calibration step of the framework. The hedge ratio $h$ determines how the insurance liabilities offsets annuity liabilities, and calibrating $h$ is therefore a critical step that directly affects the effectiveness of the natural hedge. In this section, we categorize calibration methods into three groups: optimization, immunization, and Greek neutral. Before outlining a specific technique from each category to be implemented later in this paper, we briefly review the main approaches and related literature.

    The first category is optimization, which calibrates $h$ by minimizing a chosen risk measure. \cite{cupido2024} implemented a risk-minimization approach to construct a naturally hedged portfolio that accounts for spatial dependencies between different populations. \cite{yang2019} proposed a unified hedging framework that minimizes changes in the insurer's profit function for a portfolio combining internal natural hedges and external index-based hedges. \cite{Zhu2014} applied a financial hedging approach to minimize economic capital, as originally introduced by \cite{Zhu2011}. 

    The next category is immunization, which seeks to minimize portfolio liability sensitivity to changes in mortality rates. \cite{li2012} proposed a key q-Duration framework that matches sensitivities in mortality rates to hedge longevity risk with q-forward contracts. \cite{Tsai2010} introduced a conditional Value-at-Risk minimization approach that optimizes the insurer's product mix and compared it with the modified duration matching method of \cite{HR_immunization_theory}. \cite{Chen2024} generalized the approach of \cite{Tsai2010} to incorporate both insurance supply and demand, while \cite{Gatzert2014} extended the method of \cite{HR_immunization_theory} to the entire insurance perspective, building on the framework of \cite{Gatzert2012}.
        
    Lastly, Greek neutral techniques aim to reduce portfolio liabilities with respect to changes in model-specific mortality quantities. \cite{cairns2013} introduced delta-nuga hedging as a robust approach to address recalibration risk in portfolios with index-based hedging instruments. \cite{liu2017} expanded the delta-nuga method of \cite{cairns2013} through a generalized state-space hedging framework. To hedge changes in the reserves of a naturally hedged portfolio, \cite{Luciano2017} developed delta-gamma hedging based on first- and second-order approximations. \cite{Jevti2015} applied this approach to assess the solvency of a naturally hedged portfolio from an asset-liability perspective.

    We remark that the above review is not intended to be exhaustive but rather to provide context for the calibration step of the natural hedging framework. A simple benchmark is the uncalibrated hedge, where the hedge ratio is set to $h=1$ and the insurer makes no adjustment to the insurance liabilities. Lastly, under a stochastic mortality model, the hedge ratio will need be calculated using Monte Carlo simulations. In the remainder of this section, we focus on one representative technique from each category, which will later be used in our numerical implementation.

\subsubsection{Optimization} \label{sec:FrameworkCalibrationVarMin}

    The aim of optimization is to calibrate the hedge ratio $h$ by minimizing a chosen risk measure of the hedged position. Formally, this optimization problem can be written as
    \begin{equation*}
        \min_{h \in \mathbb{R}} \rho(\mcP(h)),
    \end{equation*}
    where $\rho$ denotes a selected risk measure and $\mcP(h) = \mcA + h \cdot \mcI$ is the hedged portfolio. 
    
    When $\rho$ is taken to be variance, the problem reduces to variance minimization, a well-established approach in longevity risk management. In this case, \cite{longevity_hedge_effective_decomp} showed that the optimal hedge ratio $h^{(VM)}$ has a closed-form solution given by
    \begin{equation} \label{eq: hr var min}
        h^{(VM)} = -\frac{\text{Cov}(\mathcal{A},\mathcal{I})}{\text{Var}(\mathcal{I})}.
    \end{equation}
    The corresponding hedged portfolio is denoted by $\mcP(h^{(VM)}) = \mathcal{A} + h^{(VM)} \cdot \mcI$. This case will be used in the numerical illustrations later in the paper.

\subsubsection{Immunization}\label{sec:FrameworkCalibrationImmuniz}

    Immunization was originally introduced by \cite{redington_1952} for hedging interest rate risk. \cite{HR_immunization_theory} adapted it for natural hedging, where the goal is to neutralize the sensitivity of annuity and insurance liabilities to changes in mortality. The mortality duration of the combined portfolio $P(h)$ with respect to the force of mortality $\mu$ is defined as
    \begin{equation*}
        D^{P}_{\mu} = \frac{dP(h)}{d\mu} = \frac{dA}{d\mu} + \frac{dL(h)}{d\mu} = \frac{dA}{d\mu} + h \frac{dI}{d\mu}.
    \end{equation*}
    The optimal hedge ratio under this approach is obtained by setting $D^{P}_{\mu} = 0$.
    
    To approximate the mortality durations in $D^{P}_{\mu}$, we define the modified durations of the annuity and insurance portfolios as $D^{A}_{\mu} = \frac{A^{+} - A^{-}}{2\epsilon}$ and $D^{I}_{\mu} = \frac{I^{+} - I^{-}}{2\epsilon}$, where $\epsilon$ is a small constant, $A^{\pm}$ and $I^{\pm}$ are the expected present values of the annuity and insurance liabilities, respectively, evaluated under adjusted mortality $\mu \pm \epsilon$. These quantities capture how much the expected present values change in response to a small shift in mortality. The resulting hedge ratio is
    \begin{align} \label{eq: hr_mod_dur_match}
        h^{(DM)} = - \frac{D^{A}_{\mu}}{D^{I}_{\mu}}.
    \end{align}
    The corresponding hedged portfolio is $P(h^{(DM)}) = A + h^{(DM)}I$. This portfolio matches the modified durations of the annuity and insurance portfolios, thereby aligning their sensitivities to mortality fluctuations.

\subsubsection{Greek neutral} \label{sec:FrameworkCalibrationDeltaNeut}

    Greek neutral methods extend the notion of Greeks to mortality and longevity hedging. \cite{luciano2012delta} first introduced this idea, and \cite{longevity_greeks} further developed it for index-based longevity hedging, where the longevity delta was defined as a sensitivity measure with respect to the period effect of a stochastic mortality model. We now adapt this approach to natural hedging.  
    
    Consider the Lee-Carter (LC) model \citep{lc_model}:  
    \[
        \ln(m_{x,t}) = \alpha_{x} + \beta_{x}\kappa_t,
    \]  
    where $m_{x,t}$ is the central death rate for age $x$ in year $t$, and $\alpha_{x}$, $\beta_{x}$, and $\kappa_t$ are the LC parameters, with $\kappa_t$ following a random walk with drift. The longevity delta measures the first-order sensitivity of the survival probability $S_x(T)$ to changes in the current period effect $\kappa_0$, and is defined as $\Delta_x(T) := \frac{\partial S_x(T)}{\partial \kappa_0}$. Assuming a constant force of mortality between integer ages, \cite{longevity_greeks} showed that $\Delta_x(T)$ can be written as  
    \[
        \Delta_{x}(T) = - \sum_{s = 1}^{T} \beta_{x+s-1} \, \mathbb{E}\!\left[ \exp\!\left(Y_{x}(s) - Z_{x}(T)\right) \right],
    \]  
    where $Y_{x}(s) = \alpha_{x+s-1} + \beta_{x+s-1}\kappa_s$ and $Z_{x}(T) = \sum_{s=1}^{T} \exp(Y_{x}(s))$, with $\Delta_x(0)=0$ since $S_x(0)=1$.
    
    For the natural hedging framework, the longevity deltas of the annuity and insurance portfolios from Section \ref{sec:FrameworkValuation} can then be derived as  
    $\Delta^A = \sum_{j=1}^{n_A} \omega_j^A \sum_{k=\tau_j}^{\tau_j+t_j} c_j e^{-\delta k} \Delta_{x_j}(k)$  
    and  
    $\Delta^I = \sum_{j=1}^{n_I} \omega_j^I \sum_{k=0}^{t_j-1} b_j e^{-\delta(k+1)} \left(\Delta_{x_j}(k) - \Delta_{x_j}(k+1)\right)$.  
    A delta-neutral natural hedge would require $\Delta^A + h \Delta^I = 0$, and thus the hedge ratio is  
    \begin{align} \label{eq: hr_delta_neut_LC_model}
        h^{(DN)} = -\frac{\Delta^A}{\Delta^I}.
    \end{align}  
    The corresponding delta-neutral hedged portfolio is $P(h^{(DN)}) = A + h^{(DN)}I$. Lastly, we note that longevity delta is model-dependent. While we have illustrated it under the LC model, the delta-neutral approach can be applied to other stochastic mortality models by deriving the corresponding longevity deltas.

\subsection{Evaluation} \label{sec:FrameworkEvaluation}

    The last step of the proposed natural hedging framework is evaluation. In this step, the insurer evaluates the effectiveness of a hedged position $\mcP(h)$ using both numerical and graphical assessment tools. We discuss both types of assessments in this subsection and consider them in the toy example shown in the next subsection.

    A numerical risk measure for natural hedging can be defined in general as a mapping
    \[
        \rho: \mcP(h) \mapsto \mathbb{R},
    \]
    which assigns a real-valued assessment to the hedged position $\mcP(h)$. A fundamental example is the portfolio variance, given by $\text{Var}(\mcP(h)) = \text{Var}(\mathcal{A}) + h^{2}\text{Var}(\mathcal{I}) + 2h\,\text{Cov}(\mathcal{A},\mathcal{I})$. If the insurer focuses on downside risk, the Value-at-Risk (VaR) can be considered, defined as $\text{VaR}_{\alpha}(\mcP(h)) = \inf \{x \in \mathbb{R} : F_{\mcP(h)}(x) \geq \alpha \}$ at confidence level $\alpha \in (0,1)$, where $F_{\mcP(h)}$ is the distribution function of $\mcP(h)$. A related measure is Expected Shortfall (ES), given by $\text{ES}_{\alpha}(\mcP(h)) = \frac{1}{1-\alpha} \int_{\alpha}^{1} \text{VaR}_{u}(\mathcal{P}(h)) du$, which captures the average loss in the tail beyond VaR. For a natural hedge, these numerical risk measures can provide complementary perspectives on the risk profile of the hedged position.

    We make two remarks on numerical risk measures. First, beyond variance, VaR, and ES, other measures may also be considered, and they can be applied to transformations of the hedged position. For example, one may evaluate hedge effectiveness using $\text{VaR}_{\alpha}(\tilde{\mcP}(h))$, where $\tilde{\mcP}(h)$ is the mean-adjusted hedged portfolio given in Table~\ref{tab:Valuation}. Second, we do not aim to identify a single numerical risk measure that is universally applicable to hedged portfolios. In optimization, the chosen objective is a natural candidate, but in immunization or Greek neutral approaches no obvious measure exists. In such cases, it is important to consider multiple numerical risk measures and complementary graphical assessments.

    In conjunction with numerical measures, graphical risk metrics can further provide visual insights into the distribution of the hedged position. A common approach is to plot histograms of $\mcP(h)$ or $\tilde{\mcP}(h)$, and overlay numerical risk measures such as VaR and ES. Other useful visualizations include empirical cumulative distribution functions, Q-Q plots, or kernel density estimates. While these graphical tools are widely used, they remain limited in the depth of analysis they provide. In Section~\ref{sec:grm}, we extend beyond such standard methods and propose a new graphical risk metric tailored for natural hedging.

    In summary, the last step of our framework incorporates both numerical and graphical risk assessments to evaluate hedged positions, aiming to provide a comprehensive view of their risk profile. However, relying solely on numerical measures can overlook important aspects of hedge performance, and simple graphical diagnostics may also be insufficient. To illustrate these issues, the next subsection presents a toy example that demonstrates the implementation of the proposed natural hedging framework and motivates the new graphical risk metric introduced in Section~\ref{sec:grm}.

\subsection{Framework summary and toy example} \label{sec:FrameworkSummary_ToyEx}

    We now summarise the proposed natural hedging framework and illustrate its implementation through a toy example. Table \ref{tab: natural_hedge_framework} provides a compact reference for the three steps of the framework. Under stochastic mortality modelling, each step requires future mortality projections. In practice, Monte Carlo simulations are used to compute present values in the valuation step, hedge ratios in the calibration step, and risk measures in the evaluation step. The mortality simulation methods need not be identical across steps, which in turn allows model uncertainty to present in the natural hedge. The mortality simulation methods considered in this paper are presented in Appendix~\ref{append:mort_generator}.
    
    \begin{table}[!htpb]
        \centering
        \begin{tabular}{@{}l | l@{}}
        \toprule
        Step & Details \\ \midrule
        \begin{tabular}[c]{@{}l@{}} Step 1: \\ Valuation \end{tabular} & 
            \begin{tabular}[c]{@{}p{0.85\textwidth}@{}} 
                Derive and calculate the present values of annuity and insurance liabilities to be included in the hedge. \vspace{1ex}
            \end{tabular} \\ 
        \begin{tabular}[c]{@{}l@{}} Step 2: \\ Calibration \end{tabular} &  
            \begin{tabular}[c]{@{}p{0.85\textwidth}@{}} 
                Determine the hedge ratio, either set to $1$ for an uncalibrated hedge or calculated under a chosen calibration method. \vspace{1ex}
            \end{tabular} \\ 
        \begin{tabular}[c]{@{}l@{}} Step 3: \\ Evaluation \end{tabular} & 
            \begin{tabular}[c]{@{}p{0.85\textwidth}@{}} 
                Assess the resulting hedged positions using numerical risk measures and graphical risk assessment tools. 
            \end{tabular} \\ 
        \midrule 
        \begin{tabular}[c]{@{}l@{}} Projected \\ Mortality \end{tabular} & 
            \begin{tabular}[c]{@{}p{0.85\textwidth}@{}} 
                Mortality scenarios are required in each of the valuation, calibration, and evaluation steps. 
            \end{tabular} \\
        \bottomrule
        \end{tabular}
        \caption{The proposed natural hedging framework.}
        \label{tab: natural_hedge_framework}
    \end{table}

    To demonstrate the implementation of the proposed framework, we now present a toy example. Consider a life insurer with an annuity portfolio consisting of a single 20-year deferred 20-term life annuity issued to a 45-year-old individual, paying \$20 annually at the beginning of each year after the deferral period. To hedge this liability, the insurer issues a 30-year term life insurance policy with a death benefit of \$250, payable at the end of the year of death, to an individual aged either 40 or 50. For both issuing ages, the hedge ratio is set to $h=1$, yielding two uncalibrated hedges. The interest rate is assumed to be $i=4\%$ (or $\delta=\ln(1.04)$).
    
    Following the framework, Step 1 derives the present values of the annuity $\mcA$ and the two insurance portfolios $\mcI_1$ and $\mcI_2$ for ages 40 and 50, respectively. Step 2 constructs the calibrated insurance portfolios $\mcL_1(1)=\mcI_1$ and $\mcL_2(1)=\mcI_2$ by setting $h_1=h_2=1$. Step 3 evaluates the hedged portfolios $\mcP_1=\mcA+\mcI_1$ and $\mcP_2=\mcA+\mcI_2$ using both numerical and graphical risk assessments. Mortality scenarios are generated using the bootstrapping method described in Appendix~\ref{append:bootstrap_sim}, based on U.S. male mortality data (ages 40-99, years 1970-2018) obtained from the Human Mortality Database.\footnote{HMD. Human Mortality Database. Max Planck Institute for Demographic Research (Germany), University of California, Berkeley (USA), and French Institute for Demographic Studies (France). Available at www.mortality.org (data downloaded on 2 January 2024).}

    \autoref{tab:toy_ex} reports the mean, variance, $\text{VaR}_{0.95}$, and the excess of $\text{VaR}_{0.95}$ over the mean for the annuity portfolio $\mcA$, the two insurance portfolios $\mcL_1$ and $\mcL_2$, and the combined hedged portfolios $\mcP_1$ and $\mcP_2$. The variance is greatly reduced when either $\mcL_1$ or $\mcL_2$ is added to $\mcA$, reflecting the hedging effect of combining portfolios. Between the two hedged portfolios, $\mcP_2$ has the smaller variance and excess VaR, while $\mcP_1$ achieves the lower mean and lower VaR. The mixed results underscore the complexity of assessing hedge effectiveness in natural hedging and motivate the need for further graphical risk assessments.
    
    \begin{table}[!ht]
    \centering
    \begin{tabular}{l | c c c c}
    \toprule
    Portfolio & Mean & Variance & $\text{VaR}_{0.95}$ & $\text{VaR}_{0.95} - \text{Mean}$ \\
    \midrule
    $\mcA$ & 98.81 & 1.61 & 100.84 & 2.03 \\
    $\mcL_1$ & 22.00 & 1.01 & 23.69 & 1.68 \\
    $\mcL_2$ & 43.52 & 2.67 & 46.27 & 2.75 \\ 
    $\mcP_{1}$ & 120.81 & 0.39 & 121.83 & 1.01 \\
    $\mcP_{2}$ & 142.33 & 0.27 & 143.19 & 0.86 \\
    \bottomrule
    \end{tabular}
    \caption{Numerical risk measures for the annuity portfolio $\mcA$, the two insurance portfolios $\mcL_1$ and $\mcL_2$, and the hedged portfolios $\mcP_1$ and $\mcP_2$.}
    \label{tab:toy_ex}
    \end{table}
    
\begin{figure}[htpb]
        \centering
        \begin{subfigure}[t]{0.49\textwidth}
            \centering
            \includegraphics[width=\textwidth]{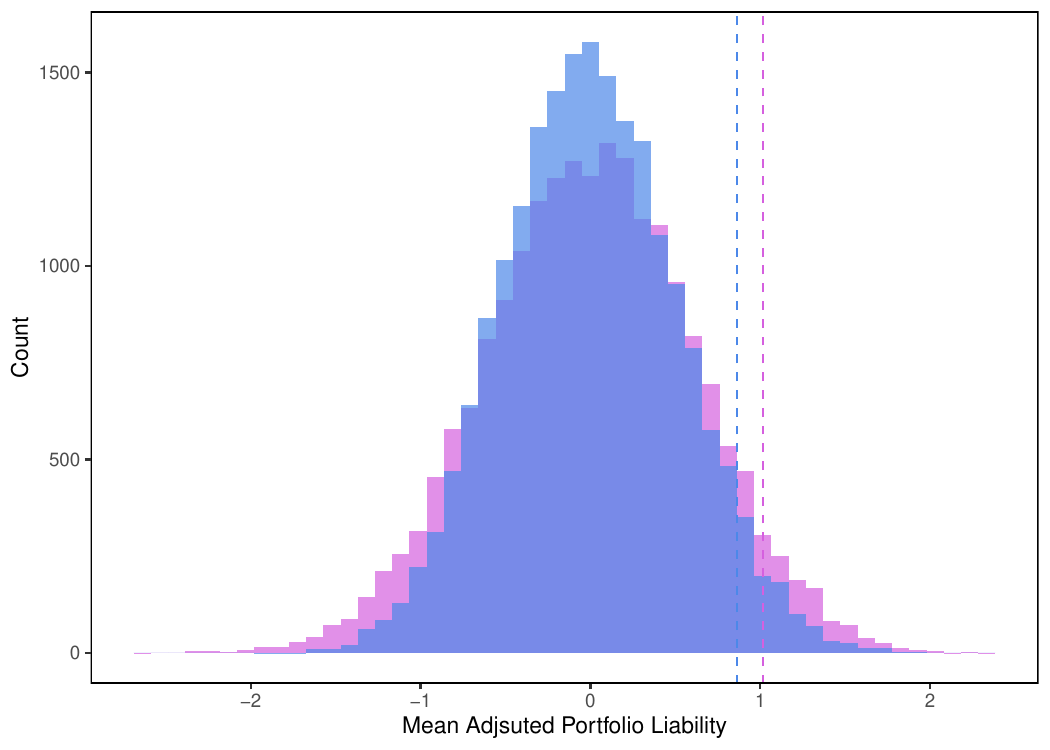}
            \caption{Liabilities for each hedged portfolio with pink and blue dashed lines denoting the $\text{VaR}_{0.95}$ for $\mcP_{1}$ (pink) and $\mcP_{2}$ (blue), respectively.}
            \label{fig: toy_ex_graph_total}
        \end{subfigure}
        \hfill
        \begin{subfigure}[t]{0.49\textwidth}
            \centering
            \includegraphics[width=\textwidth]{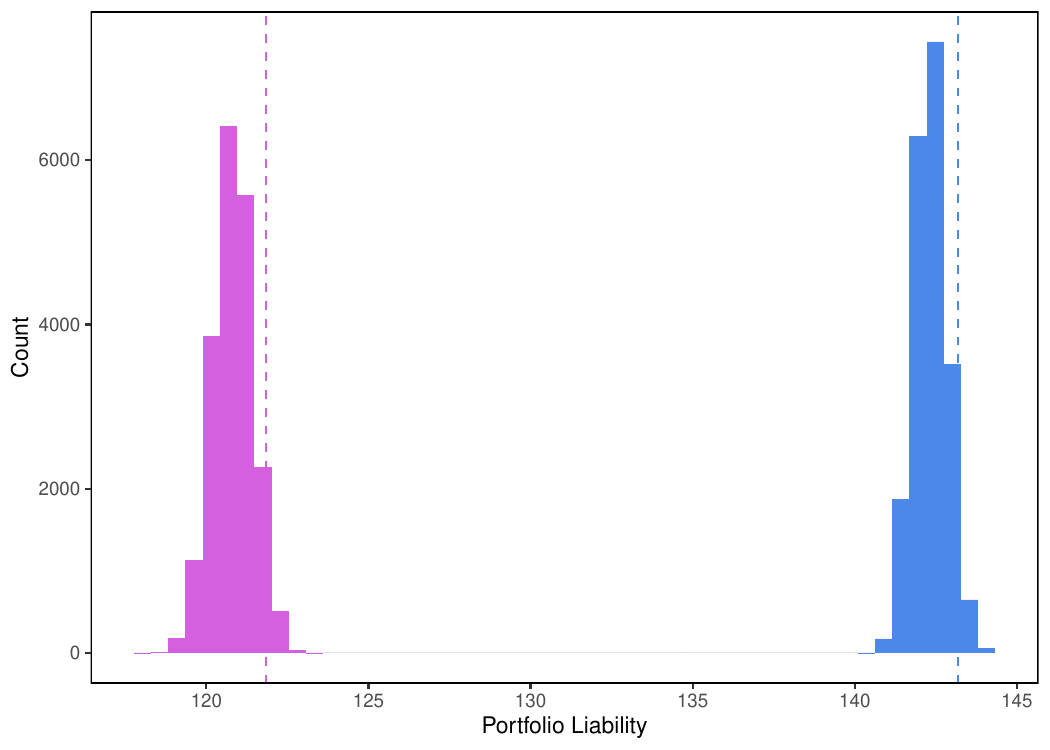}
            \caption{Mean-adjusted liabilities for each hedged portfolio with pink and blue dashed lines denoting the $\text{VaR}_{0.95}$ for $\tilde{\mcP}_{1}$ and $\tilde{\mcP}_{2}$, respectively.}
            \label{fig: toy_ex_graph_mean_adj}
        \end{subfigure}    
        \caption{Graphical risk assessment for the hedged portfolios $\mcP_{1}$ (pink) and $\mcP_{2}$ (blue).}
        \label{fig:toy_ex}
    \end{figure}

    \autoref{fig:toy_ex} provides a graphical assessment of the hedged portfolios using histograms. Panel~(a) shows the empirical distributions of $\mcP_{1}$ and $\mcP_{2}$, with dashed lines indicating their $\text{VaR}_{0.95}$. Consistent with the numerical results, $\mcP_{1}$ has a smaller $\text{VaR}_{0.95}$, reflecting lower liabilities compared to $\mcP_{2}$. Panel~(b) displays the empirical distributions of the mean-adjusted portfolios $\tilde{\mcP}_{1}$ and $\tilde{\mcP}_{2}$ (defined in Table \ref{tab:Valuation}), again with dashed lines marking their $\text{VaR}_{0.95}$. In this panel, $\mcP_{2}$ shows smaller deviations from its mean than $\mcP_{1}$, confirming the lower variance observed in the numerical measures.

    This toy example illustrates how different numerical and graphical risk assessments can lead to inconclusive results about hedge effectiveness. More importantly, such assessments cannot capture the underlying relationship between the annuity portfolio $\mcA$ and the insurance portfolio $\mcL$ within the hedged position $\mcP$. For example, when the realized liability of $\mcP$ exceeds its $\text{VaR}_{0.95}$, it is unclear whether this outcome reflects weak offsetting effect between $\mcA$ and $\mcL$, or whether both portfolios simultaneously contribute to large losses. Answering such questions can help insurers to make more informed decisions in portfolio selection and in the design of natural hedges, which we address in Section~\ref{sec:grm}.
   
\section{The Graphical Risk Metric} \label{sec:grm}

    This section introduces a new graphical risk metric for natural hedging. The metric provides a deeper analysis of the hedged position by examining the joint distribution of the underlying annuity and insurance portfolios on a two-dimensional scale. In the following, we first describe how this graphical risk metric is constructed and then develop an interpretation procedure that enables an enhanced risk assessment of the hedged position. To illustrate the construction and interpretation, we continue to use the toy example outlined in Section~\ref{sec:FrameworkSummary_ToyEx}.

\subsection{Constructing the metric} \label{sec:grm_construction}

    The construction of our graphical risk metric is inspired by the work of \cite{graphical_basis_risk_measure}, who introduced a visual tool for evaluating population basis risk in index-based longevity hedges. A brief review of the original method by \cite{graphical_basis_risk_measure} is provided in Appendix~\ref{append:grm_review}. We adapt this approach to the context of natural hedging and construct a two-dimensional assessment of the joint distribution of the underlying annuity and insurance portfolios, rather than focusing solely on the hedged position. This representation enables direct visualisation of the dependence structure between the annuity and insurance liabilities.

    \begin{figure}[!ht]
        \centering
        \includegraphics[width=0.6\textwidth]{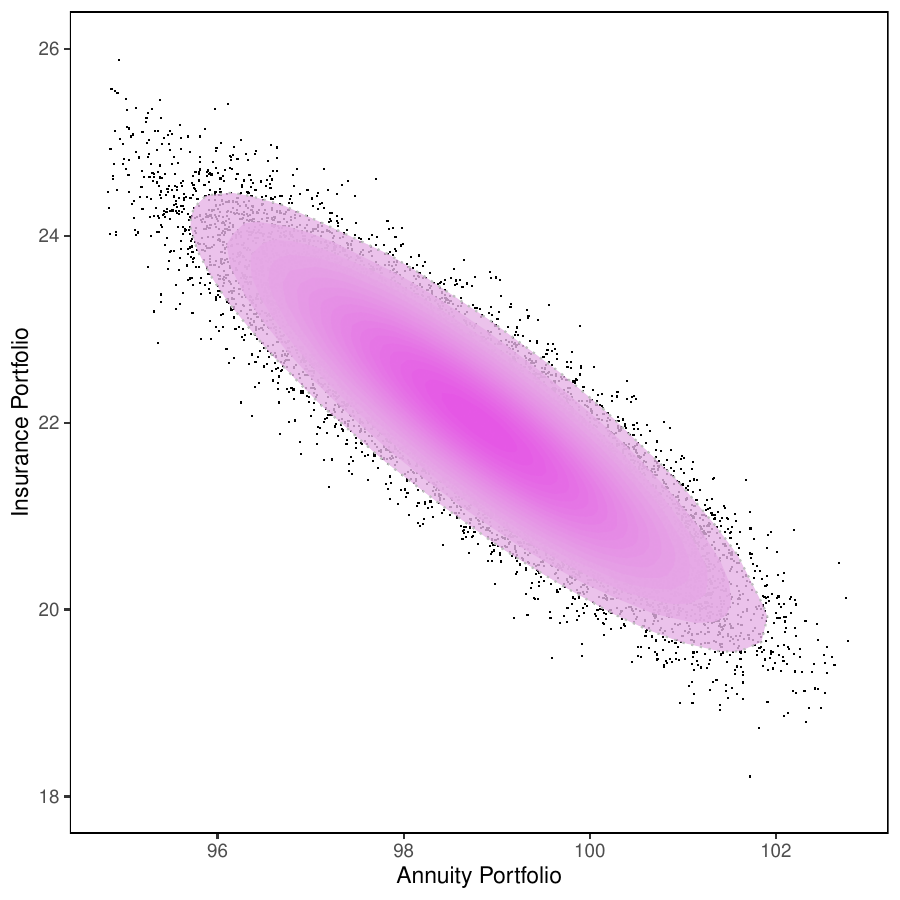}
         \caption{Graphical risk metric illustrating the joint distribution of the annuity portfolio $\mcA$ and the insurance portfolio $\mcL_1$ for the hedged position $\mcP_{1}$ from the toy example.}
        \label{fig:grm_one_port}
    \end{figure}

    Figure~\ref{fig:grm_one_port} illustrates the result of applying the metric to the hedged position $\mcP_{1}$ from the toy example. Details on how this figure is constructed are provided in Appendix~\ref{append:grm_construction}. The x- and y-axes represent the present values of the annuity portfolio $\mcA$ and the insurance portfolio $\mcL_1$, respectively. Each black dot corresponds to a realisation of $(\mcA, \mcL_1)$, while the shaded pink regions denote joint prediction regions at different confidence levels. The figure reveals a clear inverse relationship between $\mcA$ and $\mcL_1$, forming a downward-sloping pattern in both the point cloud and the shaded regions.

\begin{figure}[!ht]
        \centering
        \begin{subfigure}[b]{0.475\textwidth}
            \centering
            \includegraphics[width=\textwidth]{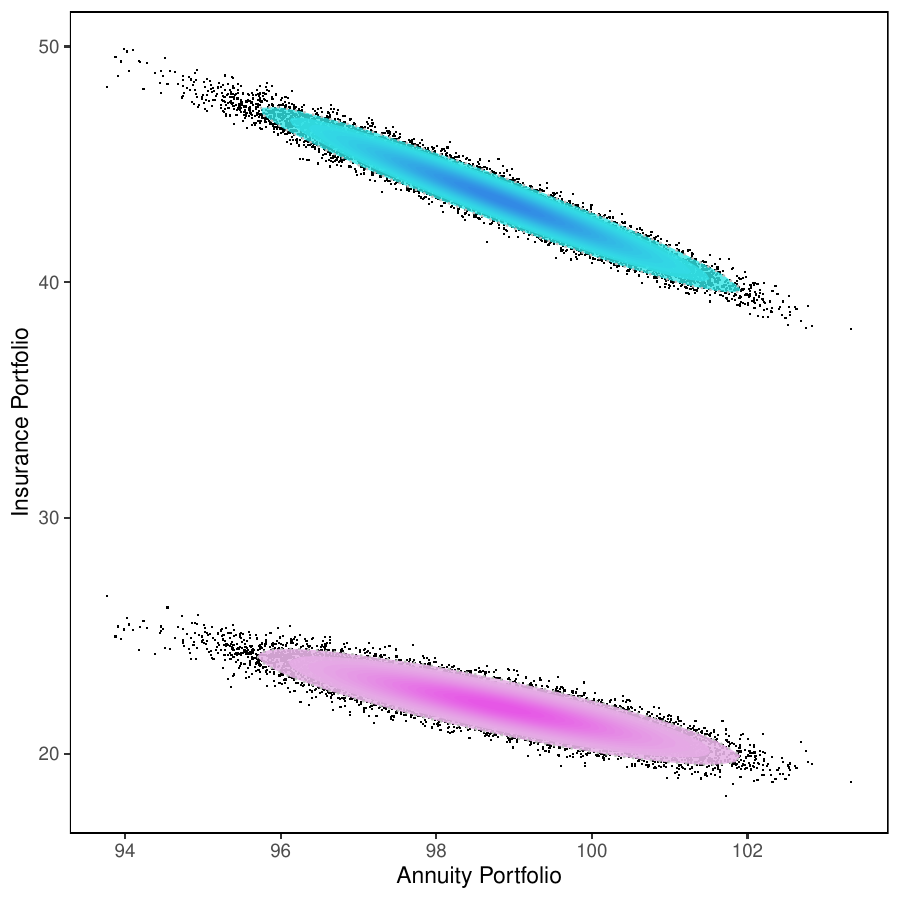}
            \caption{Unadjusted}
            \label{fig:sub_two_portfolios_no_adjust}
        \end{subfigure}
        \hfill
        \begin{subfigure}[b]{0.475\textwidth}
            \centering
            \includegraphics[width=\textwidth]{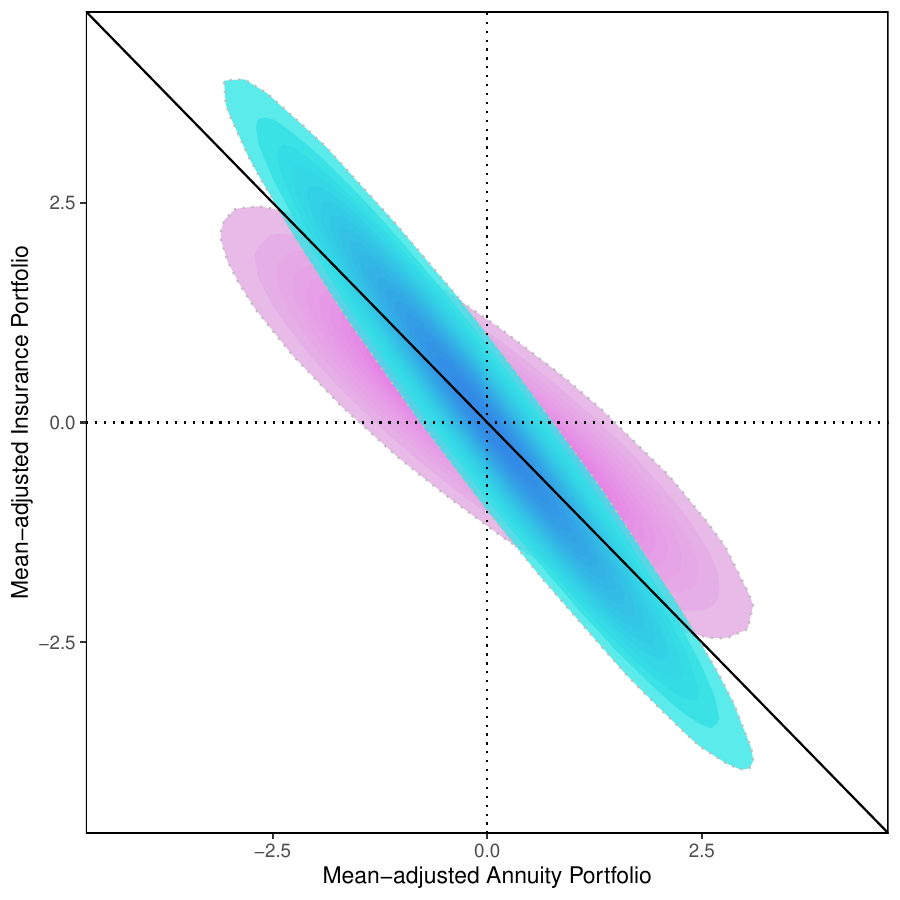}
            \caption{Mean-adjusted}
            \label{fig:sub_two_portfolios_centered}
        \end{subfigure}
        \caption{Graphical risk metric illustrating the effectiveness of the hedged positions $\mcP_{1}$ (pink) and $\mcP_{2}$ (blue) from the toy example.}
        \label{fig:grm_two_ports}
    \end{figure}

    The graphical risk metric illustrated in Figure~\ref{fig:grm_one_port} enables a visual evaluation of the hedge effectiveness of a single portfolio. The same construction can be extended to compare multiple hedged portfolios, which is particularly useful when several hedging opportunities are available. Figure~\ref{fig:grm_two_ports} illustrates such an extension by jointly displaying the annuity portfolio $\mcA$ against two insurance portfolios, $\mcL_1$ and $\mcL_2$, corresponding to the hedged positions $\mcP_1$ and $\mcP_2$ from the toy example. This extension allows for a direct comparison of alternative natural hedges under a single visual platform.

    Panel~(a) of Figure~\ref{fig:grm_two_ports} plots the annuity portfolio $\mcA$ against the corresponding insurance portfolios $\mcL_{1}$ and $\mcL_{2}$. Since the annuity is identical across both portfolios, the larger insurance liabilities of $\mcL_{2}$ (relative to $\mcL_{1}$) result in higher total liabilities, consistent with the numerical measures reported in \autoref{tab:toy_ex}. This outcome is expected, as the life insurance policy from $\mcP_{2}$ is issued to a 50-year-old individual, whose higher mortality risk will lead to greater liabilities than the 40-year-old policy from $\mcP_{1}$.

    To enable a fairer comparison, Panel~(b) of Figure~\ref{fig:grm_two_ports} presents the mean-adjusted annuity and insurance liabilities, represented by $(\tilde{\mcA}, \tilde{\mcL}_{1})$ and $(\tilde{\mcA}, \tilde{\mcL}_{2})$, whose joint prediction regions are centred at the origin. The solid black line with slope $-1$ represents the \textit{benchmark} line, corresponding to scenarios in which deviations in the annuity and insurance liabilities perfectly offset (i.e., $\tilde{\mcA} + \tilde{\mcL} = 0$), resulting in a ``perfect'' hedge. Building on this benchmark, we next introduce an interpretative procedure to analyse potential hedging outcomes based on the mean-adjusted portfolios.

\subsection{Interpreting the metric} \label{sec:grm_interpretation}

    To gain a comprehensive visual understanding of the risk profile underlying a hedged position, we now develop an interpretation aid for the proposed graphical risk metric. This aid functions as a diagnostic tool for analysing the joint behaviour of the annuity and insurance portfolios. By representing all present value realisations in the $(\tilde{\mcA}, \tilde{\mcL})$ plane, it enables visual assessment of how effectively deviations in the insurance liability offset those in the annuity liability. 

    \begin{figure}[!ht]
        \centering
        \begin{subfigure}[b]{0.475\textwidth}
            \centering
            \includegraphics[width=\textwidth]{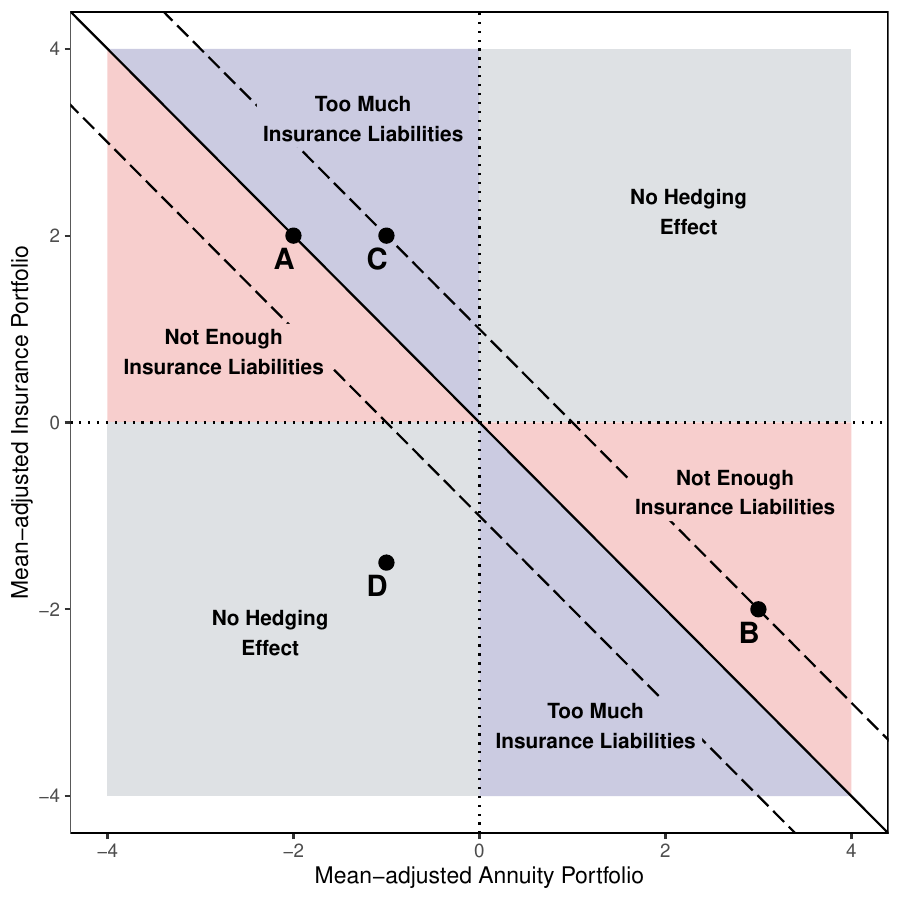}
            \caption{Interpretation}
            \label{fig:grm_interpret}
        \end{subfigure}
        \hfill
        \begin{subfigure}[b]{0.475\textwidth}
            \centering
            \includegraphics[width=\textwidth]{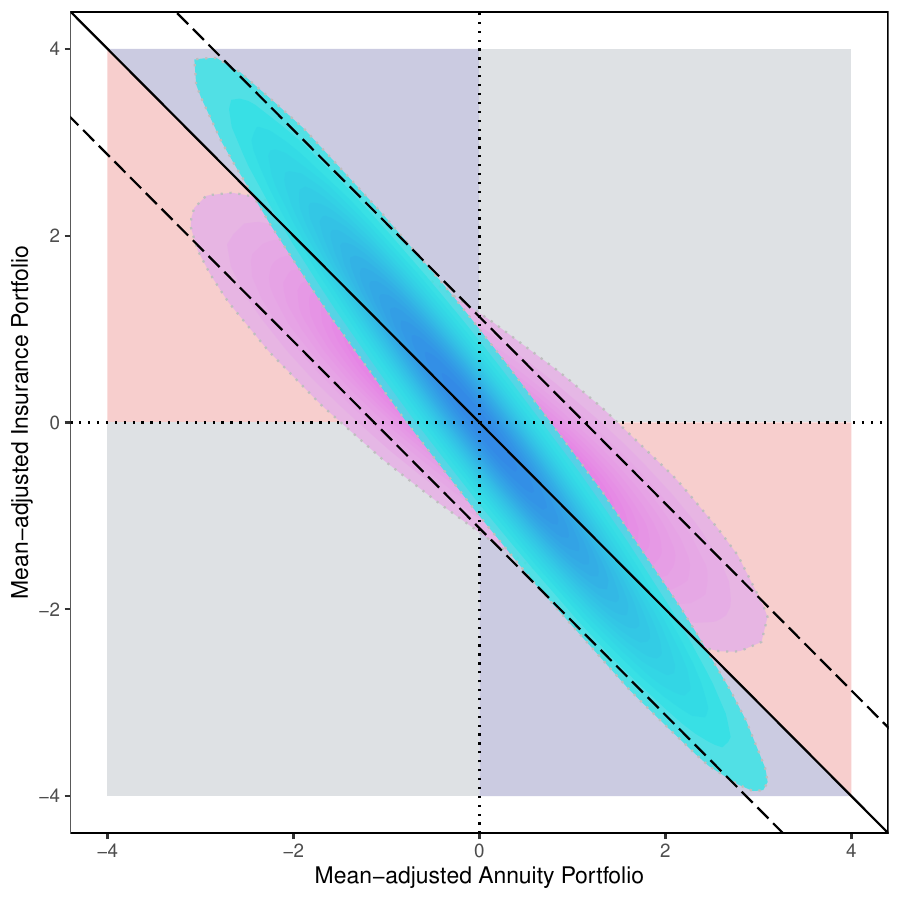}%
            \caption{Application}
            \label{fig:grm_method_compare}
        \end{subfigure}
        \caption{Interpretation procedure for the proposed graphical risk metric and its application to the hedged positions $\mcP_{1}$ (pink) and $\mcP_{2}$ (blue) from the toy example.}
        \label{fig:grm_method}
    \end{figure}
    
    Figure~\ref{fig:grm_method} illustrates our interpretation procedure. Panel~(a) defines the shaded regions and benchmark lines used to classify hedging outcomes, where the shading corresponds to different types of hedging outcome and the dashed lines indicate the magnitude of deviation from a perfect hedge. To clarify the interpretation of these shaded regions, we provide brief explanations of the scenarios marked by Points A, B, C, and D in Figure~\ref{fig:grm_interpret}:  
    \begin{itemize}
        \item \textbf{Point A}: The annuity liability is \$2 below its expected value, while the insurance liability is \$2 above. The surplus in the insurance liability fully offsets the annuity shortfall, representing a perfectly hedged scenario. All points along the benchmark line correspond to this ideal outcome.  
        \item \textbf{Point B}: The annuity liability is \$3 above its expectation, but the insurance liability is only \$2 below. The shortfall in the insurance liability leaves a net liability surplus of \$1 for the hedged position, corresponding to the case of \emph{Not Enough Insurance Liabilities} (red region).  
        \item \textbf{Point C}: The annuity liability is \$1 below its expected value, while the insurance liability is \$2 above. The excess in the insurance liability leads to a net liability surplus of \$1 for the hedged position, representing the case of \emph{Too Much Insurance Liabilities} (blue region). 
        \item \textbf{Point D}: Both the annuity and insurance liabilities fall below their expectations, by \$1 and \$1.5 respectively. This scenario creates a combined liability deficit of \$2.5, indicating the case of \emph{No Hedging Effect} (grey region), where losses occur simultaneously in both portfolios.
    \end{itemize}
    
\begin{figure}[!ht]
        \centering
        \begin{subfigure}[b]{0.475\textwidth}
            \centering
            \includegraphics[width=\textwidth]{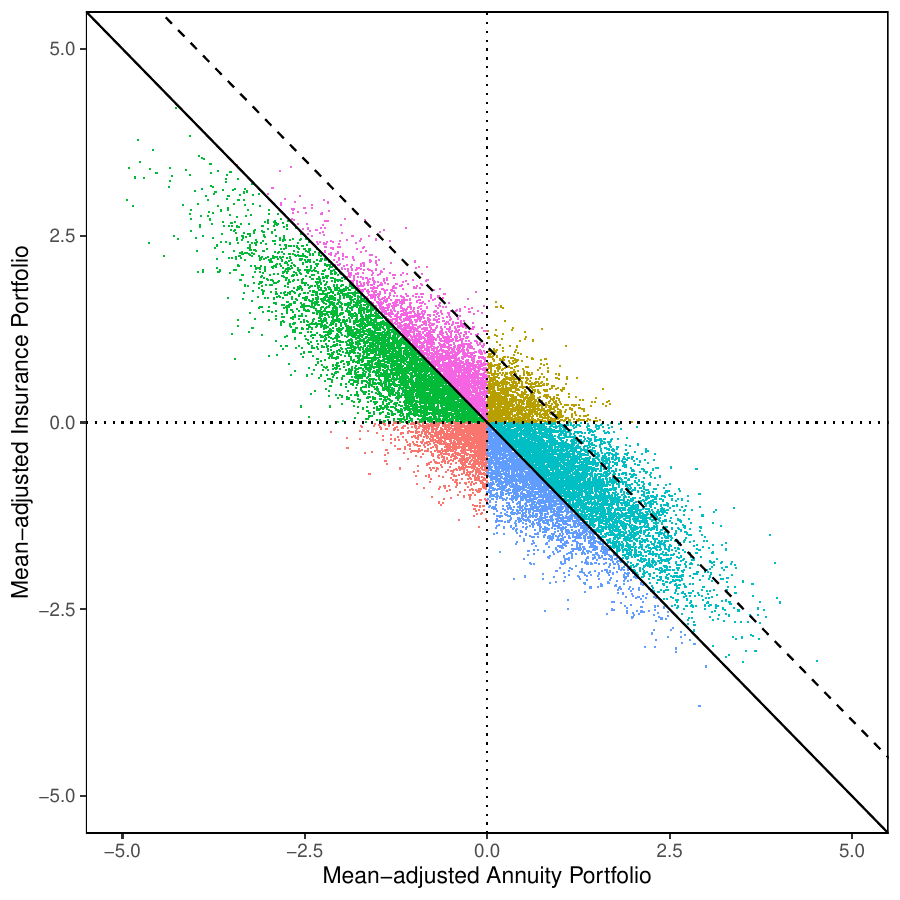}
            \caption{Hedged position $\tilde{\mcP}_{1}$}
            \label{fig:grm_outcome_port1}
        \end{subfigure}
        \hfill
        \begin{subfigure}[b]{0.475\textwidth}
            \centering
            \includegraphics[width=\textwidth]{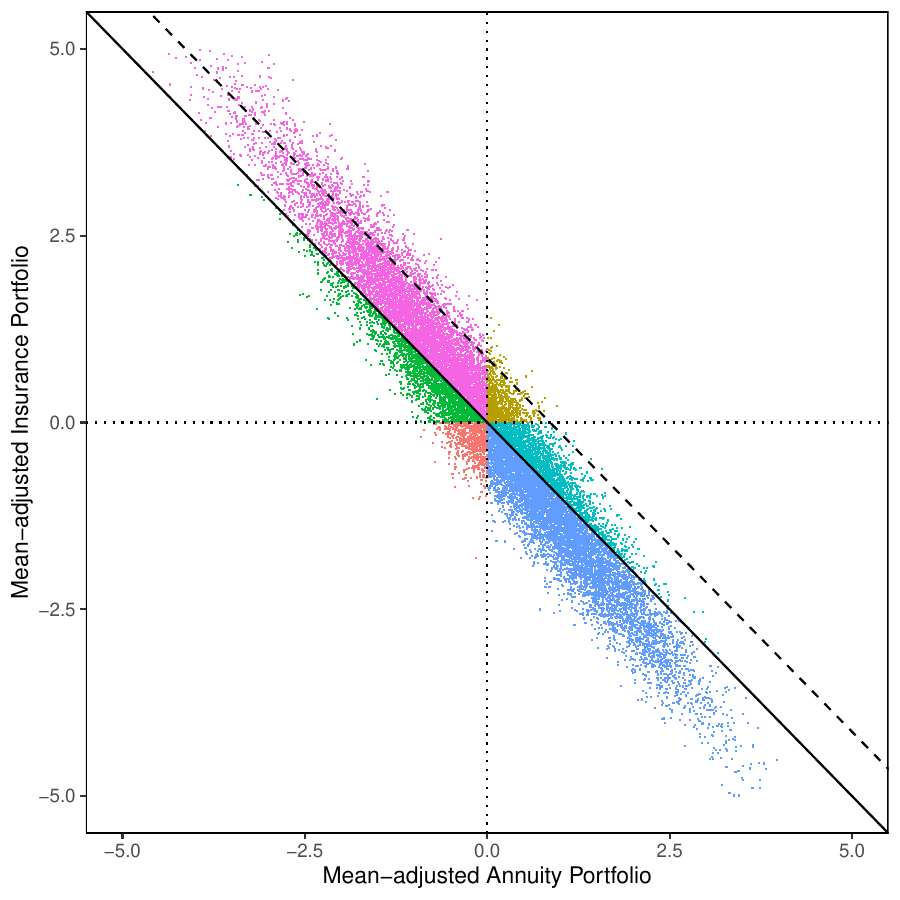}
            \caption{Hedged position $\tilde{\mcP}_{2}$}
            \label{fig:grm_outcome_port2}
        \end{subfigure}
        \caption{Simulated outcomes of $\tilde{\mcP}_{1}$ and $\tilde{\mcP}_{2}$, coloured by hedging outcome with the dashed line denoting the mean-adjusted $\VaR_{0.95}$.}
        \label{fig:grm_outcome}
    \end{figure}

    The two dashed lines in Figure~\ref{fig:grm_interpret} represent equal magnitudes of deviation from the expected total liability, corresponding to either a surplus or a deficit of \$1. Since Point D lies below the lower dashed line, it immediately means that it has a larger magnitude of deviation than Points B and C. Although Points B and C both have a surplus of \$1, they arise from different causes; Point B has insufficient insurance liabilities, while Point C has excessive insurance liabilities. This distinction would not be apparent in a one-dimensional graphical assessment, such as a histogram. Figure~\ref{fig:grm_outcome} extends this idea by showing simulated realisations coloured by hedging outcome types.
  
    Figure~\ref{fig:grm_outcome} presents 20,000 simulated realisations of $(\tilde{\mcA}, \tilde{\mcL}_{1})$ and $(\tilde{\mcA}, \tilde{\mcL}_{2})$ in Panels (a) and (b), respectively. Points are coloured by the type of hedging outcome. Pink and blue indicate \emph{Too Much Insurance Liabilities}, leading to a surplus and deficit in total liability, respectively. Green and teal denote \emph{Not Enough Insurance Liabilities}, leading to a deficit and surplus in total liability, respectively. Red and gold represent \emph{No Hedging Effect}, where both the insurance and annuity portfolios deviate in the same direction. The dashed lines mark the mean-adjusted $\text{VaR}_{0.95}$ of the corresponding hedged positions. Any realisations above these lines exceed the respective $\text{VaR}_{0.95}$ values reported in \autoref{tab:toy_ex}.

    Comparing the simulated hedging outcomes in Figure~\ref{fig:grm_outcome}, $\tilde{\mcP}_{1}$ shows more realisations associated with \emph{Not Enough Insurance Liabilities} (green and teal) and \emph{No Hedging Effect} (red and gold) than $\tilde{\mcP}_{2}$. Conversely, $\tilde{\mcP}_{2}$ exhibits a higher frequency of \emph{Too Much Insurance Liabilities} (pink and blue), indicating that its insurance exposure tends to overcompensate for annuity deviations. Among the outcomes that exceed the mean-adjusted $\text{VaR}_{0.95}$ as marked by the dashed lines, those from $\tilde{\mcP}_{1}$ primarily arise from \emph{Not Enough Insurance Liabilities} (teal) and \emph{No Hedging Effect} (gold), while those from $\tilde{\mcP}_{2}$ are mostly driven by \emph{Too Much Insurance Liabilities} (pink).

    We now return to Figure~\ref{fig:grm_method_compare}, which applies the interpretation aid to compare $\tilde{\mcP}_{1}$ and $\tilde{\mcP}_{2}$ by overlaying their joint prediction regions. The dashed lines above and below the solid benchmark line represent equal magnitudes of surplus and deficit in the hedged position. It is clear that the joint prediction region of $\tilde{\mcP}_{1}$ extends further into the area associated with \emph{Not Enough Insurance Liabilities}, while $\tilde{\mcP}_{2}$ covers more of the \emph{Too Much Insurance Liabilities} area. More importantly, the joint prediction region of $\tilde{\mcP}_{1}$ extends above the upper dashed line, indicating more severe (liability surplus) scenarios than that of $\tilde{\mcP}_{2}$.

    In conclusion, the proposed graphical risk metric complements numerical evaluation by exposing the underlying hedging behaviour. Specifically, $\mcP_{2}$ achieves stronger offsets between annuity and insurance liabilities but is more prone to excess insurance exposure, whereas $\mcP_{1}$ faces greater downside risk from insufficient offsetting. These findings highlight how the graphical risk metric reveals important behavioural differences between hedged positions. In the next section, we apply this graphical framework to a broader set of numerical illustrations to further explore its practical use and interpretive value.

\section{Numerical Illustrations} \label{sec:illustrations}

    We now demonstrate how the natural hedging framework and the graphical risk metric can be applied to address practical natural hedging problems. Three illustrations are considered: (1) identifying the most effective insurance portfolio, (2) selecting the optimal hedge calibration technique, and (3) evaluating model risk arising from alternative mortality projections. Each illustration follows the same framework as in Section~\ref{sec:framework} and employs the graphical risk metric developed in Section~\ref{sec:grm}. Appendix~\ref{append:ill_weights} provides the details of the annuity and insurance portfolios considered in this section.

\subsection{Illustration 1: Insurance Portfolio Selection} \label{subsec:ill1_insur_port}

    Our first illustration investigates how to identify an effective insurance portfolio for natural hedging. Consider an insurer managing an annuity portfolio issued to individuals aged 40-60, with \$10,000 annual payments starting at age 65. The insurer considers three candidate insurance portfolios, each providing a death benefit of \$750,000 payable at the end of the year of death: (1) whole life policies issued to individuals aged 40-60 ($\mcI_{1}$), (2) whole life policies issued to individuals aged 40-49 ($\mcI_{2}$), and (3) 20-year term policies issued to individuals aged 40-49 ($\mcI_{3}$). The annual interest rate is assumed to be 4\%, and the limiting age is set at 100. Table~\ref{tab:ill1_framework} summarises how the natural hedging framework is applied to this illustration. The three steps are applied consistently, with the only difference arising from the features of the insurance portfolios.

    \begin{table}[!htpb]
        \centering
        \begin{tabular}{@{}l | l@{}}
        \toprule
        Step & Details \\ \midrule
        \begin{tabular}[c]{@{}l@{}} Step 1: \\ Valuation \end{tabular} & 
            \begin{tabular}[c]{@{}p{0.85\textwidth}@{}} 
                Derive the present values of the annuity portfolio (issuing ages 40-60, deferred to age 65, 35 annual payments of \$10,000) and three candidate insurance portfolios with a \$750,000 death benefit: $\mcI_{1}$ (issuing ages 40-60, whole life), $\mcI_{2}$ (issuing ages 40-49, whole life), and $\mcI_{3}$ (issuing ages 40-49, 20-year term).
            \end{tabular} \\ 
        \begin{tabular}[c]{@{}l@{}} Step 2: \\ Calibration \end{tabular} &  
            \begin{tabular}[c]{@{}p{0.85\textwidth}@{}}  
                Calibrate hedge ratios $h_{1}$, $h_{2}$, and $h_{3}$ using the variance-minimisation method.
            \end{tabular} \\ 
        \begin{tabular}[c]{@{}l@{}} Step 3: \\ Evaluation \end{tabular} & 
            \begin{tabular}[c]{@{}p{0.85\textwidth}@{}} 
                Construct the hedged portfolios $\mcP_{1}$, $\mcP_{2}$, and $\mcP_{3}$, and evaluate them comparatively using both numerical risk measures and the proposed graphical risk metric.
            \end{tabular} \\ 
        \midrule 
        \begin{tabular}[c]{@{}l@{}} Projected \\ Mortality \end{tabular} & 
            \begin{tabular}[c]{@{}p{0.85\textwidth}@{}}
                Mortality scenarios are generated using the bootstrapping method described in Appendix \ref{append:bootstrap_sim}, applied consistently across the valuation, calibration, and evaluation steps.
            \end{tabular} \\
        \bottomrule
        \end{tabular}
        \caption{Natural hedging framework applied to Illustration 1.}
        \label{tab:ill1_framework}
    \end{table}

\begin{figure}[!htpb]
        \centering
        \begin{subfigure}[b]{0.475\textwidth}
            \centering
            \includegraphics[width=\textwidth]{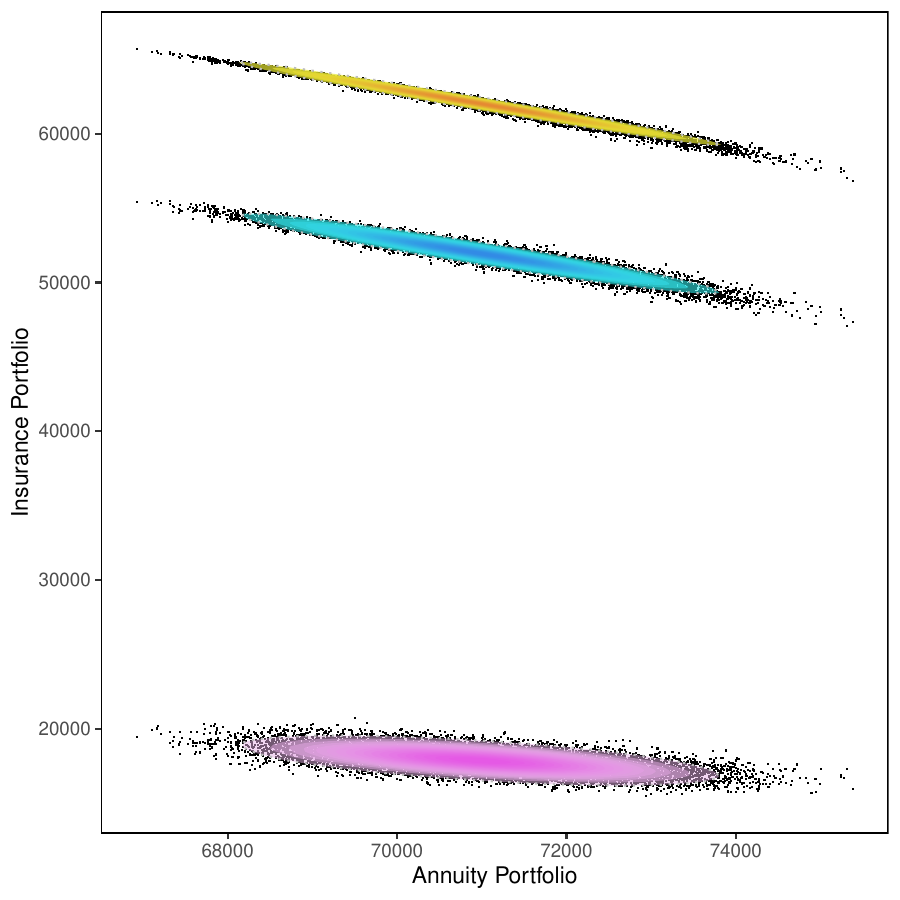}
            \caption{Unadjusted}
            \label{fig:ill1_grm_unadj}
        \end{subfigure}
        \hfill
        \begin{subfigure}[b]{0.475\textwidth}
            \centering
            \includegraphics[width=\textwidth]{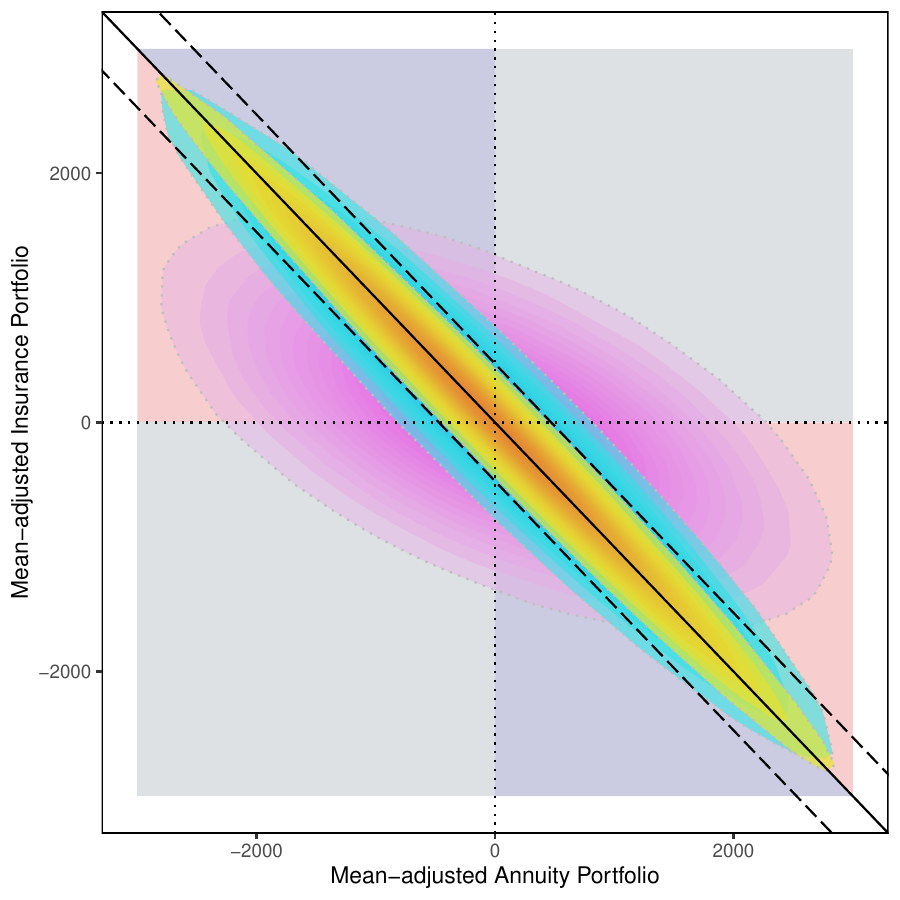}
            \caption{Mean-adjusted}
            \label{fig:ill1_grm_adj}
        \end{subfigure}
        \caption{Graphical risk metric applied to the hedged positions $\mcP_{1}$ (yellow), $\mcP_{2}$ (blue) and $\mcP_{3}$ (pink) from Illustration 1.}
        \label{fig:ill1_grm}
    \end{figure}

    Table~\ref{tab:ill1_results} reports the calibrated hedge ratios and numerical risk measures for the three hedged portfolios. Based on the variance and the mean-adjusted $\VaR_{0.95}$, $\mcP_{1}$ outperforms $\mcP_{2}$ and $\mcP_{3}$, whereas the $\VaR_{0.95}$ indicates that $\mcP_{3}$ performs best. To complement these numerical results and reveal the underlying hedging behaviour, Figure~\ref{fig:ill1_grm} presents the graphical risk metric for the three hedged portfolios. In Panel~(a), the unadjusted version shows that the three portfolios differ in the magnitude of their calibrated insurance liabilities, with $\mcP_{1}$ having the largest and $\mcP_{3}$ the smallest. The mean-adjusted version in Panel~(b) shows that $\tilde{\mcP}_{1}$ aligns most closely with the benchmark line, indicating stronger offsetting between annuity and insurance liabilities, $\tilde{\mcP}_{2}$ exhibits greater variability with more outcomes falling into the \emph{Not Enough Insurance Liabilities} and \emph{Too Much Insurance Liabilities} regions, and $\tilde{\mcP}_{3}$ displays the widest dispersion and more frequent \emph{No Hedging Effect}.
    
    \begin{table}[!htpb]
        \centering
        \begin{tabular}{@{}l | c c c c@{}}
        \toprule
        Portfolio & Hedge Ratio & Variance & $\VaR_{0.95}$ & Mean-adjusted $\VaR_{0.95}$ \\ 
        \midrule
        $\mcP_{1} = \mcA + h_{1}\cdot\mcI_{1}$ & 0.267 & 34,090 & 133,295 & 290 \\ 
        $\mcP_{2} = \mcA + h_{2}\cdot\mcI_{2}$ & 0.275 & 112,252 & 123,496 & 563 \\ 
        $\mcP_{3} = \mcA + h_{3}\cdot\mcI_{3}$ & 0.311 & 875,566 & 90,413 & 1,571 \\ 
        \bottomrule
        \end{tabular}
        \caption{Calibrated hedge ratios and numerical risk measures for Illustration 1.}
        \label{tab:ill1_results}
    \end{table}

    In summary, this illustration demonstrates how the proposed framework and graphical risk metric together facilitate the selection of an appropriate insurance portfolio for natural hedging. Consistent with expectations, $\mcP_{1}$ with whole life policies issued to the same age range as the annuity portfolio provides the most effective hedge. Restricting the insurance portfolio to younger issuing ages, as in $\mcP_{2}$, increases variability and weakens the offset between annuity and insurance liabilities. Further shortening the policy term to 20 years, as in $\mcP_{3}$, amplifies this effect and leads to more outcomes with \emph{No Hedging Effect}. These results highlight that when issuing ages or policy terms deviate from those of the underlying annuity portfolio, the effectiveness of natural hedging reduces and the residual risk becomes more pronounced.

\subsection{Illustration 2: Hedge Calibration} \label{subsec:ill2_hedge_calib}

    The second illustration examines how different hedge calibration methods affect the effectiveness of a natural hedge. Consider an insurer managing an annuity portfolio issued to individuals aged 40-60, with \$10,000 annual payments starting at age 65 for 20 years. The insurer constructs an insurance portfolio of whole life policies issued to the same age range and applies three different calibration methods for determining the hedge ratio: variance-minimisation (VM), duration-matching (DM), and delta-neutral (DN). Table~\ref{tab:ill2_framework} summarises how the natural hedging framework is applied to this illustration. The valuation and evaluation steps remain identical across the three methods, with the only difference lying in the calibration step.

    \begin{table}[!ht]
        \centering
        \begin{tabular}{@{}l | l@{}}
        \toprule
        Step & Details \\ \midrule
        \begin{tabular}[c]{@{}l@{}} Step 1: \\ Valuation \end{tabular} & 
            \begin{tabular}[c]{@{}p{0.85\textwidth}@{}} 
                Derive the present values of the annuity portfolio (issuing ages 40-60, deferred to 65, 20 annual payments of \$10,000) and the insurance portfolio of whole life policies with a \$750,000 death benefit issued to individuals aged 40-60. \vspace{1ex}
            \end{tabular} \\ 
        \begin{tabular}[c]{@{}l@{}} Step 2: \\ Calibration \end{tabular} &  
            \begin{tabular}[c]{@{}p{0.85\textwidth}@{}}  
                Calibrate the hedge ratio using three different methods: $h^{(VM)}$ from the variance-minimisation method, $h^{(DM)}$ from the duration-matching method, and $h^{(DN)}$ from the delta-neutral method. \vspace{1ex}
            \end{tabular} \\ 
        \begin{tabular}[c]{@{}l@{}} Step 3: \\ Evaluation \end{tabular} & 
            \begin{tabular}[c]{@{}p{0.85\textwidth}@{}} 
                Construct the hedged portfolios $\mcP^{(VM)}$, $\mcP^{(DM)}$, and $\mcP^{(DN)}$, and evaluate them comparatively using both numerical risk measures and the proposed graphical risk metric.
            \end{tabular} \\ 
        \midrule 
        \begin{tabular}[c]{@{}l@{}} Projected \\ Mortality \end{tabular} & 
            \begin{tabular}[c]{@{}p{0.85\textwidth}@{}}
                Mortality scenarios are generated using the Lee-Carter model described in Appendix \ref{append:mort_model_sim}, applied consistently across the valuation, calibration, and evaluation steps.
            \end{tabular} \\
        \bottomrule
        \end{tabular}
        \caption{Natural hedging framework applied to Illustration 2.}
        \label{tab:ill2_framework}
    \end{table}
    
\begin{figure}[!ht]
        \centering
        \begin{subfigure}[b]{0.475\textwidth}
            \centering
            \includegraphics[width = \textwidth]{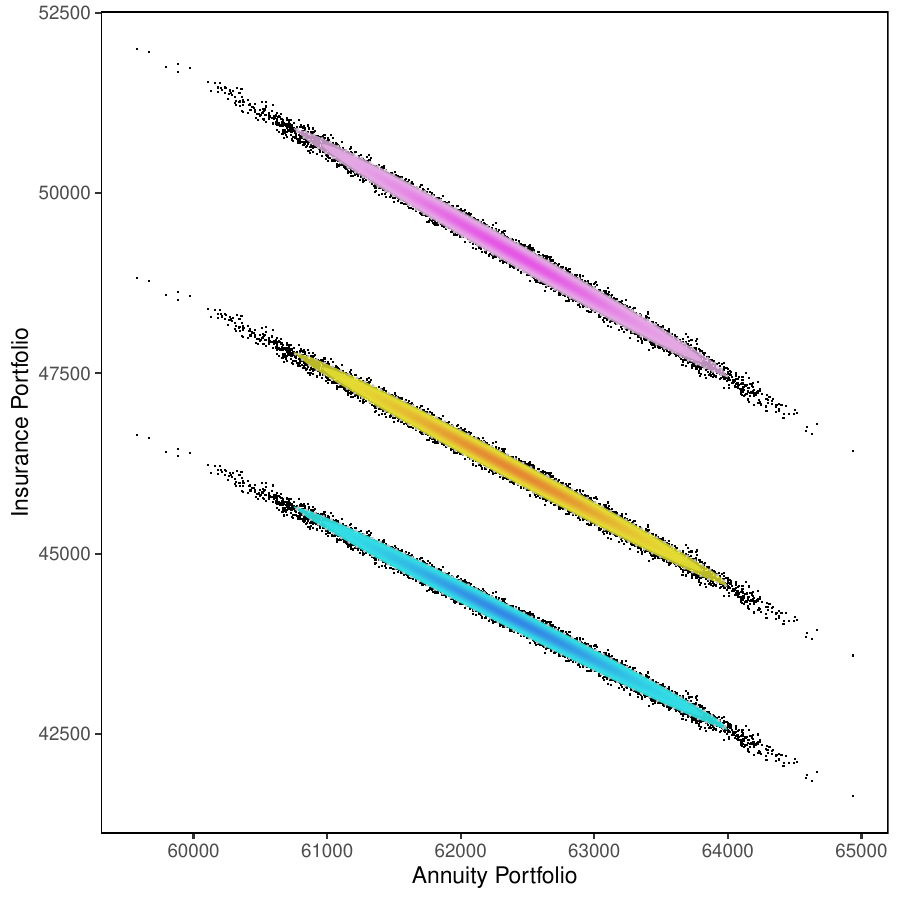}
            \caption{Unadjusted}
            \label{fig:ill2_grm_unadj}
        \end{subfigure}
        \hfill
        \begin{subfigure}[b]{0.475\textwidth}
            \centering
            \includegraphics[width=\textwidth]{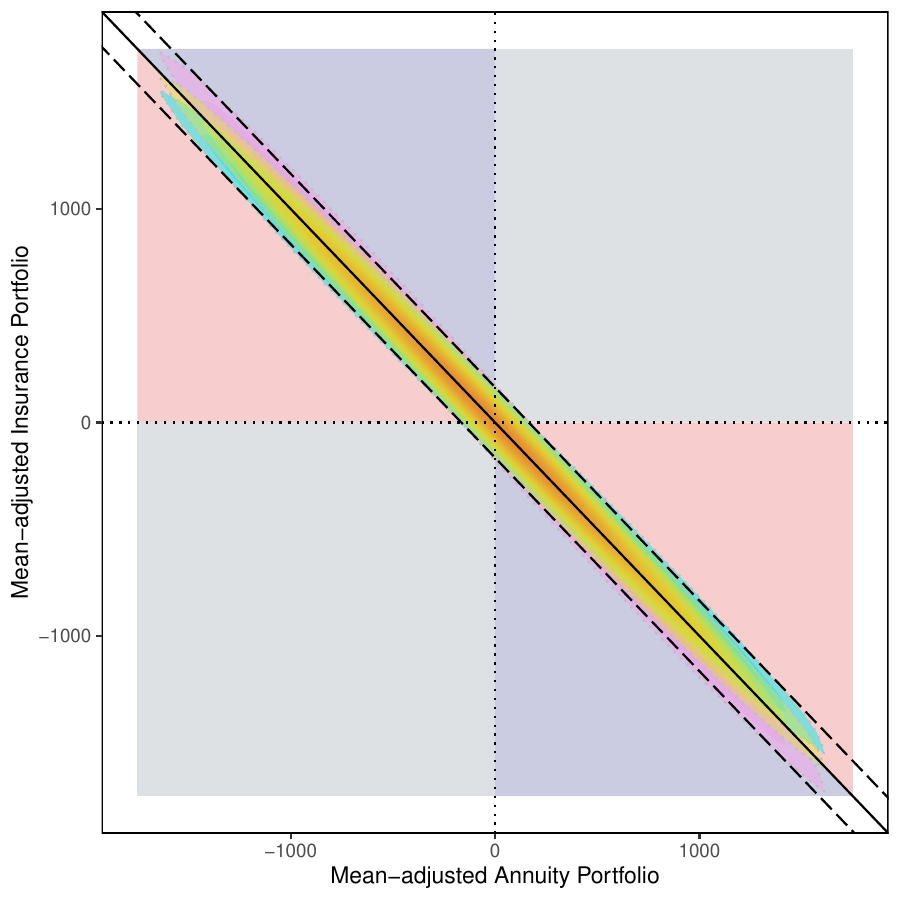}
            \caption{Mean-adjusted}
            \label{fig:ill2_grm_adj}
        \end{subfigure}
        \caption{Graphical risk metric applied to the hedged positions $\mcP^{(VM)}$ (yellow), $\mcP^{(DM)}$ (blue), and $\mcP^{(DN)}$ (pink) from Illustration 2.}
        \label{fig:ill2_grm}
    \end{figure}

    Table~\ref{tab:ill2_results} reports the calibrated hedge ratios and numerical risk measures for the three portfolios. The hedge ratios range between 0.189 and 0.211, with the smallest from duration matching and the largest from delta neutral. As expected, $\mcP^{(VM)}$ achieves the lowest variance while $\mcP^{(DN)}$ exhibits the highest risk across all measures. Figure~\ref{fig:ill2_grm} compares the three hedged portfolios using the graphical risk metric. In Panel~(a), the unadjusted portfolios show that $\mcP^{(DM)}$ requires the least amount of insurance liabilities, while $\mcP^{(DN)}$ requires the most, consistent with their respective hedge ratios. Panel~(b) displays the mean-adjusted portfolios, where $\tilde{\mcP}^{(VM)}$ lies closest to the benchmark line, indicating balanced offsetting between annuity and insurance liabilities. In contrast, $\tilde{\mcP}^{(DM)}$ tilts toward the \emph{Not Enough Insurance Liabilities} region, and $\tilde{\mcP}^{(DN)}$ toward the \emph{Too Much Insurance Liabilities} region, revealing systematic under- and over-hedging tendencies.

    \begin{table}[!ht]
        \centering
        \begin{tabular}{@{}l | c c c c@{}}
        \toprule
        Portfolio & Hedge Ratio & Variance & $\VaR_{0.95}$ & Mean-adjusted $\VaR_{0.95}$ \\ 
        \midrule
        $\mcP^{(VM)} = \mcA + h^{(VM)}\cdot\mcI$ & 0.198 & 4,816 & 108,670 & 113 \\ 
        $\mcP^{(DM)} = \mcA + h^{(DM)}\cdot\mcI$ & 0.189 & 5,744 & 106,614 & 123 \\ 
        $\mcP^{(DN)} = \mcA + h^{(DN)}\cdot\mcI$ & 0.211 & 6,642 & 111,690 & 132 \\ 
        \bottomrule
        \end{tabular}
        \caption{Calibrated hedge ratios and numerical risk measures for Illustration 2.}
        \label{tab:ill2_results}
    \end{table}

    In summary, this illustration shows how the choice of hedge ratio calibration method influences both numerical performance and hedging behaviour. The variance-minimisation method achieves the most balanced hedge, with the lowest variance and a symmetric distribution of hedging outcomes. The duration-matching approach tends to under-hedge, while the delta-neutral approach tends to over-hedge, leaving asymmetric residual risk exposure in the hedged position. The numerical and graphical results jointly provide a coherent view of the trade-offs underlying different calibration methods.

\subsection{Illustration 3: Model Risk} \label{subsec:ill3_model_risk}

    Our final illustration examines the impact of model risk on natural hedging effectiveness. Suppose an insurer uses the Lee-Carter (LC) model to calibrate a natural hedge but is concerned that actual mortality may not be accurately described by it. To investigate the risk that the mortality model assumed for calibration differs from the one generating actual future mortality, the insurer considers two additional models for evaluation: the Cairns-Blake-Dowd (CBD) model and a non-parametric bootstrapping (BS) model. The goal is to assess whether the hedged position constructed by the LC model is robust to alternative mortality generating methods. Table~\ref{tab:ill3_framework} summarises how the natural hedging framework is applied to this setting. The annual interest rate is again assumed to be 4\%, and the limiting age is set at 100.

    \begin{table}[!ht]
        \centering
        \begin{tabular}{@{}l | l@{}}
        \toprule
        Step & Details \\ \midrule
        \begin{tabular}[c]{@{}l@{}} Step 1: \\ Valuation \end{tabular} & 
            \begin{tabular}[c]{@{}p{0.85\textwidth}@{}} 
                Derive the present values of the annuity portfolio (issuing ages 40-60, deferred to age 65, 35 annual payments of \$10,000) and the whole-life insurance portfolio with a \$750,000 death benefit issued to the same ages. \vspace{1ex}
            \end{tabular} \\ 
        \begin{tabular}[c]{@{}l@{}} Step 2: \\ Calibration \end{tabular} &  
            \begin{tabular}[c]{@{}p{0.85\textwidth}@{}}  
                Calibrate the hedge ratio $h$ using the variance-minimisation method with mortality scenarios generated from the LC model. \vspace{1ex}
            \end{tabular} \\ 
        \begin{tabular}[c]{@{}l@{}} Step 3: \\ Evaluation \end{tabular} & 
            \begin{tabular}[c]{@{}p{0.85\textwidth}@{}} 
                Construct and compare the hedged portfolios $\mcP^{(LC)}$, $\mcP^{(CBD)}$, and $\mcP^{(BS)}$, which are evaluated using mortality scenarios generated by the LC, CBD, and BS models, respectively.
            \end{tabular} \\ 
        \midrule 
        \begin{tabular}[c]{@{}l@{}} Projected \\ Mortality \end{tabular} & 
            \begin{tabular}[c]{@{}p{0.85\textwidth}@{}}
                Mortality scenarios are generated from the LC model in the calibration step, and from the LC, CBD, and BS models in the evaluation step. The mortality generating processes are provided in Appendix \ref{append:mort_generator}.
            \end{tabular} \\
        \bottomrule
        \end{tabular}
        \caption{Natural hedging framework applied to Illustration 3.}
        \label{tab:ill3_framework}
    \end{table}

    Table~\ref{tab:ill3_results} reports the calibrated hedge ratio and numerical risk measures for the three portfolios. The hedge ratio is identical across all portfolios ($h = 0.310$). When evaluated by the same model used in the calibration step, $\mcP^{(LC)}$ exhibits the lowest variance and mean-adjusted $\VaR_{0.95}$. However, when mortality is generated from an alternative model, $\mcP^{(CBD)}$ shows substantially higher variability, while $\mcP^{(BS)}$ lies between $\mcP^{(LC)}$ and $\mcP^{(CBD)}$ in both variance and mean-adjusted $\VaR_{0.95}$. These results suggest that when a natural hedge is both calibrated and evaluated under the LC model, its hedge effectiveness may be significantly overestimated.

    \begin{table}[!ht]
        \centering
        \begin{tabular}{@{}l | c c c c@{}}
        \toprule
        Portfolio & Hedge Ratio & Variance & $\VaR_{0.95}$ & Mean-adjusted $\VaR_{0.95}$ \\ 
        \midrule
        $\mcP^{(LC)}$ & 0.310 & 1,628 & 143,093 & 67 \\ 
        $\mcP^{(CBD)}$ & 0.310 & 245,611 & 142,538 & 697 \\ 
        $\mcP^{(BS)}$ & 0.310 & 66,711 & 143,290 & 350 \\ 
        \bottomrule
        \end{tabular}
        \caption{Calibrated hedge ratio and numerical risk measures for Illustration 3.}
        \label{tab:ill3_results}
    \end{table}

\begin{figure}[!ht]
        \centering
        \begin{subfigure}[b]{0.475\textwidth}
            \centering
            \includegraphics[width=\textwidth]{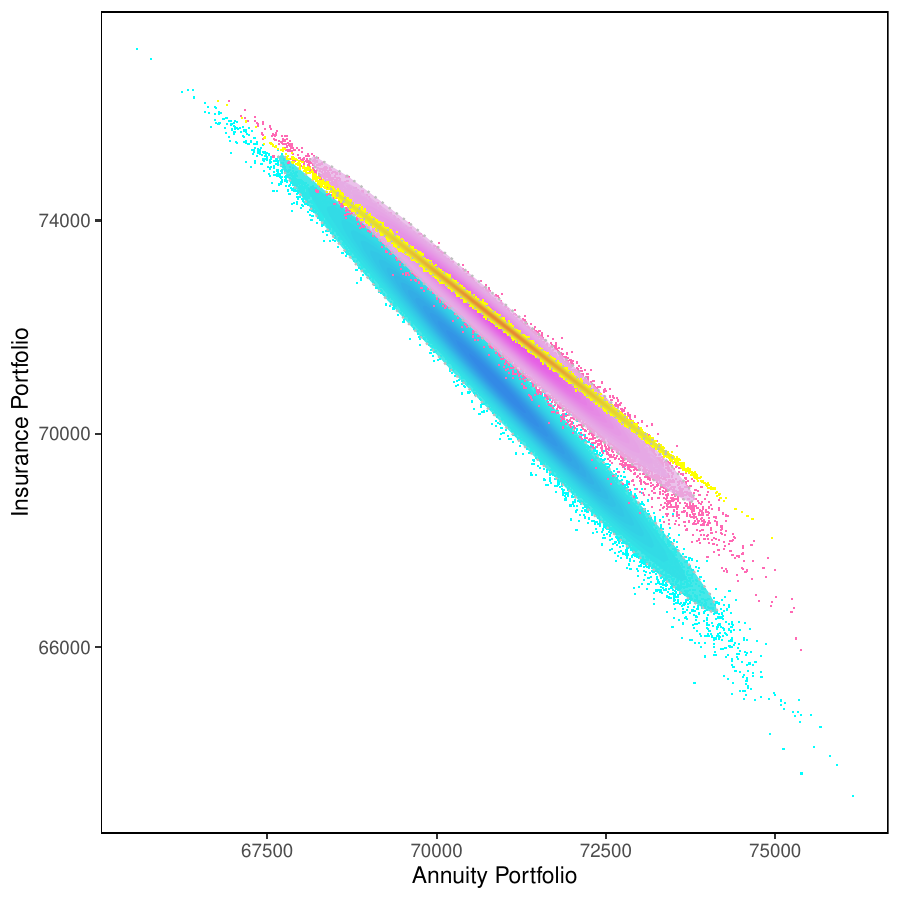}
            \caption{Unadjusted}
        \end{subfigure}
        \hfill
        \begin{subfigure}[b]{0.475\textwidth}
            \centering
            \includegraphics[width=\textwidth]{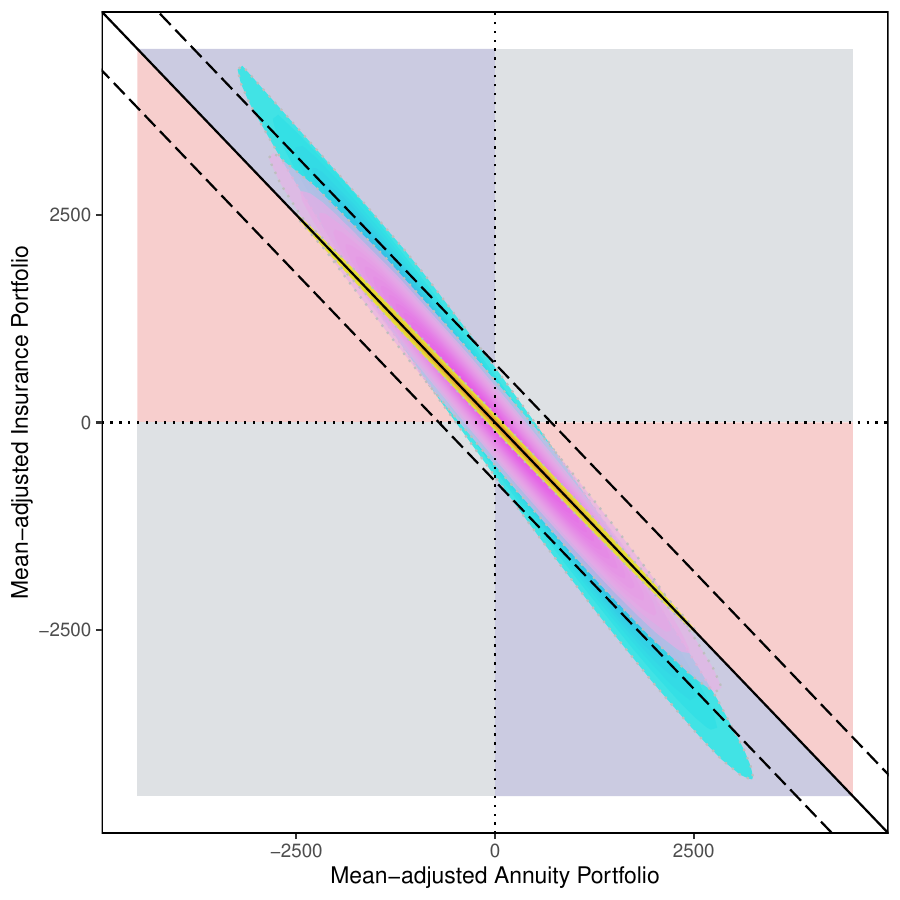}
            \caption{Mean-adjusted}
        \end{subfigure}
        \caption{Graphical risk metric applied to the hedged positions $\mcP^{(LC)}$ (yellow), $\mcP^{(CBD)}$ (blue), and $\mcP^{(BS)}$ (pink) in Illustration~3.}
        \label{fig:ill3_grm}
    \end{figure}

    Figure~\ref{fig:ill3_grm} compares the three portfolios using the graphical risk metric. In Panel~(a), the unadjusted version shows that $\mcP^{(LC)}$ exhibits the smallest joint prediction region, while $\mcP^{(CBD)}$ and $\mcP^{(BS)}$ display much wider dispersion in both the annuity and insurance portfolios. The mean-adjusted version in Panel~(b) reveals that $\tilde{\mcP}^{(LC)}$ lies closest to the benchmark line, whereas both $\tilde{\mcP}^{(CBD)}$ and $\tilde{\mcP}^{(BS)}$ tilt toward the \emph{Too Much Insurance Liabilities} region. This suggests that the LC-calibrated natural hedge tends to require an excessive amount of insurance liabilities, resulting in suboptimal hedging outcomes when actual mortality deviates from the LC model. This is an insight that cannot be easily inferred from the numerical results alone.

\begin{figure}[!ht]
        \centering
        \begin{subfigure}[b]{0.475\textwidth}
            \centering
            \includegraphics[width=\textwidth]{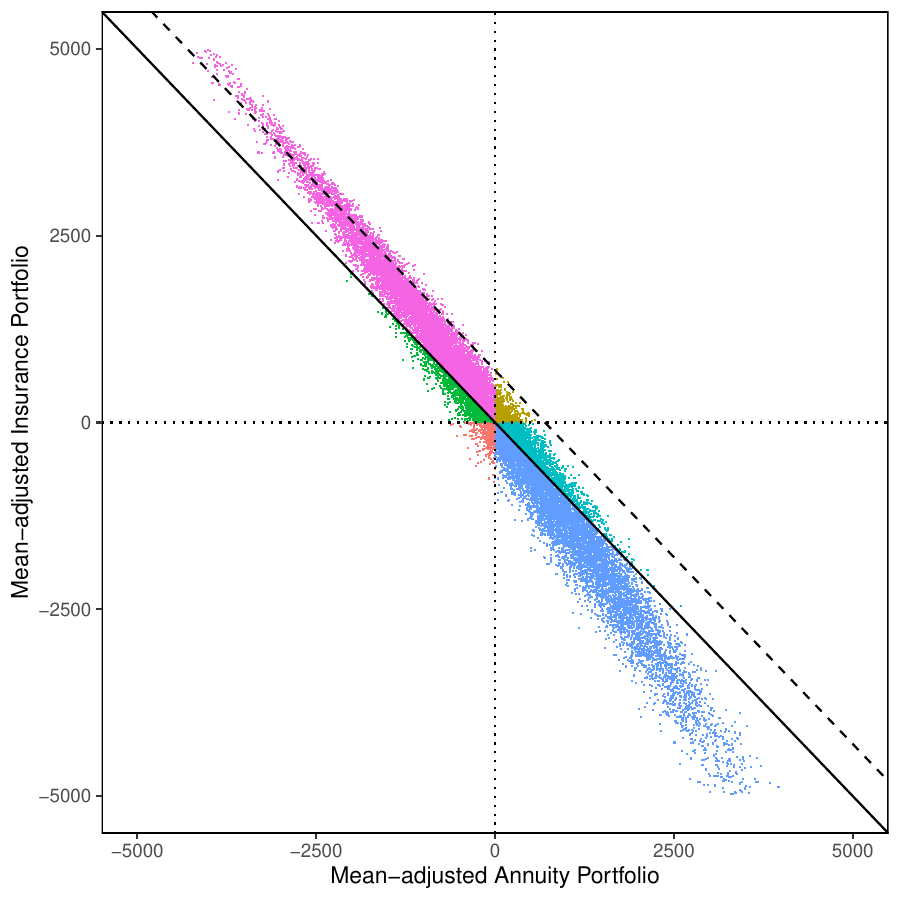}
            \caption{Hedged position $\tilde{\mcP}^{(CBD)}$}
        \end{subfigure}
        \hfill
        \begin{subfigure}[b]{0.475\textwidth}
            \centering
            \includegraphics[width=\textwidth]{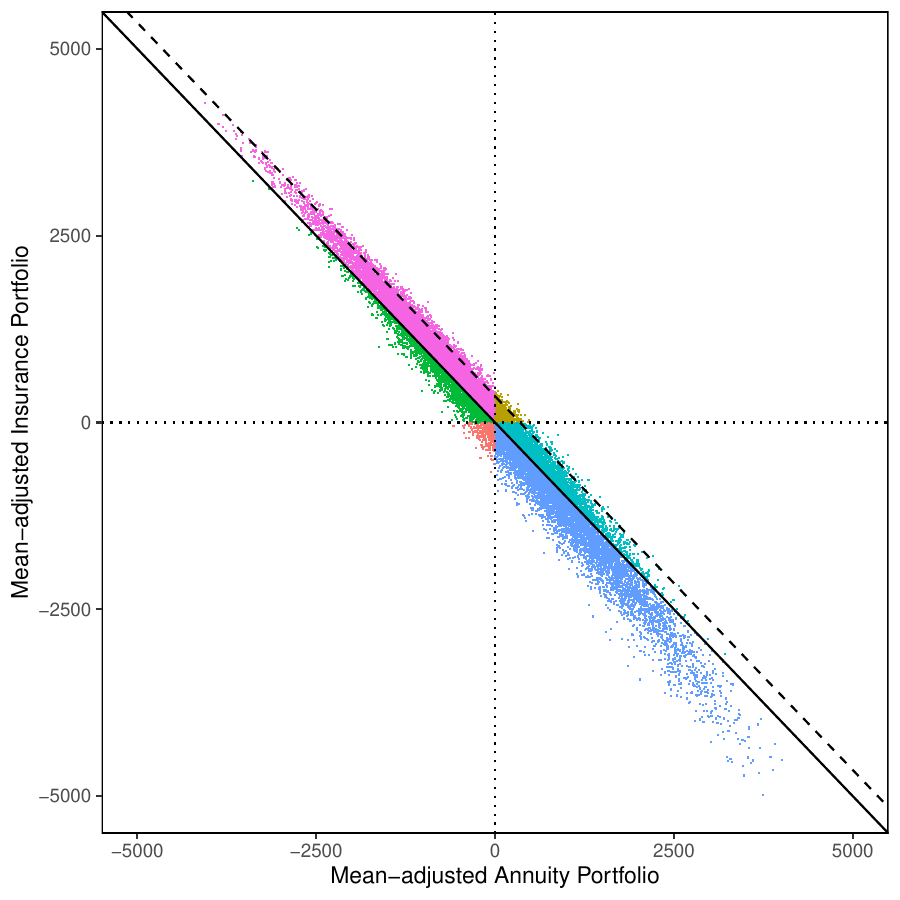}
            \caption{Hedged position $\tilde{\mcP}^{(BS)}$}
        \end{subfigure}
        \caption{Simulated outcomes of $\tilde{\mcP}^{(CBD)}$ and $\tilde{\mcP}^{(BS)}$, coloured by hedging outcome with the dashed line denoting the mean-adjusted $\VaR_{0.95}$.}
        \label{fig:ill3_outcomes}
    \end{figure}

    From Figure~\ref{fig:ill3_grm}, we see that $\tilde{\mcP}^{(CBD)}$ extends beyond the dashed lines, while $\tilde{\mcP}^{(BS)}$ remains within them, indicating that the CBD model implies more volatile and extreme hedging outcomes than the non-parametric BS approach. To further illustrate their differences, Figure~\ref{fig:ill3_outcomes} plots simulated realisations of $\tilde{\mcP}^{(CBD)}$ and $\tilde{\mcP}^{(BS)}$, coloured by hedging outcome. Both portfolios show a concentration of realisations in the \emph{Too Much Insurance Liabilities} region. However, when compared against their respective mean-adjusted $\VaR_{0.95}$ thresholds (marked by the dashed lines), the exceedances for $\tilde{\mcP}^{(CBD)}$ occur mainly due to \emph{Too Much Insurance Liabilities}, whereas for $\tilde{\mcP}^{(BS)}$, the exceedances arise across all three hedging outcome types.

    In summary, this illustration demonstrates that model risk can significantly distort the perceived effectiveness of a natural hedge. When the mortality model used for evaluation differs from the one used in calibration, the hedged position exhibits suboptimal performance and excess volatility under our natural hedging framework. The proposed graphical risk metric complements traditional numerical measures by revealing how these distortions arise from differing mortality dynamics and by exposing variations in hedging outcome types that numerical summaries alone would overlook.


\section{Conclusion} \label{sec:conclusion}

    This paper developed a comprehensive framework for constructing and evaluating natural hedging strategies, complemented by a graphical risk metric specifically designed for visual assessment of natural hedges. The proposed framework unifies the valuation, calibration, and evaluation steps of natural hedging to provide a structured process for comparing different portfolio settings, calibration techniques, and mortality scenarios. The graphical risk metric provides a new evaluation dimension by visualizing the joint distribution of annuity and insurance liabilities and by distinguishing different hedging outcome types. Together, these risk management tools form an integrated approach for analysing longevity and mortality risks underlying a life insurer’s balance sheet.

    The proposed natural hedging framework and graphical risk metric offer several practical benefits. The framework enables consistent implementation of natural hedging across multiple policy types, calibration techniques, and model assumptions. It is also highly flexible and can accommodate various existing or newly developed natural hedging strategies beyond those demonstrated in this paper. The graphical risk metric complements the framework as an innovative evaluation tool for assessing and comparing multiple natural hedges when making hedging decisions. It further enhances interpretability by revealing whether poor performance arises from insufficient or excessive diversification, or from a lack of offsetting effects between liabilities.

    Using three numerical illustrations, we demonstrated the capability and diagnostic value of the proposed methods. The results show that the framework can identify trade-offs among different insurance portfolios, assess calibration techniques across variance-minimisation, duration-matching, and delta-neutral approaches, and evaluate hedge robustness under alternative mortality scenario generators. In all cases, the graphical risk metric reveals dependencies and asymmetries in hedge performance that are otherwise hidden within numerical risk measures. Although the underlying risks differ across illustrations, the proposed methods effectively address each scenario, demonstrating their flexibility and practical applicability.

    This study has several limitations that call for further development. First, future research could extend the framework to incorporate stochastic interest rates, dynamic hedging strategies, and alternative mortality models. Second, the graphical risk metric can be enhanced by asymmetric or non-convex joint prediction regions to capture skewed risk profiles and sudden mortality shocks. Finally, the natural hedging framework could be expanded to integrate solvency capital requirements and to support intuitive visualization for regulatory solvency assessments. Together, these potential directions would further advance the integration of natural hedging within longevity risk management research.


\section*{Conflict of Interest}
The authors declare that they have no competing interests.

\section*{Data Availability}
Replication code and data used in this study are available from the corresponding author upon reasonable request. All simulated mortality and portfolio data were generated using publicly available mortality inputs from the Human Mortality Database.

\section*{Funding}
This research was supported by the \textit{Natural Sciences and Engineering Research Council of Canada} under Grant No. RGPIN-2025-04157 and DGECR-2025-00488. The funder had no role in study design, data collection and analysis, decision to publish, or preparation of the manuscript.

\bibliographystyle{apalike}
\bibliography{reference}


\appendix
\setcounter{equation}{0}
\setcounter{figure}{0}
\setcounter{table}{0}

\numberwithin{equation}{section}
\numberwithin{figure}{section}
\numberwithin{table}{section}

\renewcommand{\theequation}{\thesection.\arabic{equation}}
\renewcommand{\thefigure}{\thesection.\arabic{figure}}
\renewcommand{\thetable}{\thesection.\arabic{table}}

\section{Metric Construction} 

\subsection{Review of \cite{graphical_basis_risk_measure}} \label{append:grm_review}

    \cite{graphical_basis_risk_measure} proposed a graphical risk metric to visually assess population basis risk that cannot be transferred through an index-based longevity hedge. It serves as a visualization tool to identify the reference population with the lowest basis risk. In this appendix, we briefly review the construction of this risk metric, which we adapted for developing the graphical risk metric for natural hedging.

    Suppose a hedger seeks to mitigate the longevity risk of its mortality-dependent liabilities associated with population $H$ that is proportional to a survivor measure $S^{(H)}$. The hedger selects a mortality-dependent derivative linked to population $R$, with a payoff proportional to a survivor index $S^{(R)}$. The graphical population basis risk metric aims to visualize the potential deviations between the mortality dependent liabilities $S^{(H)}$ and the derivatives $S^{(R)}$. Let $C^{(i)} = S^{(i)} - \mathbb{E}(S^{(i)})$ be the mean-adjusted value of $S^{(i)}$ for $i = R$ and $H$. Using $C^{(i)}$ instead of $S^{(i)}$ has two advantages: it focuses on deviations from the expectation rather than the values themselves, and it centers the realisations of $S^{(i)}$ at the origin, allowing visual comparison of outcome of different reference populations $R$.
    
    Prior to constructing the graphical metric, \cite{graphical_basis_risk_measure} incorporates a variance minimization hedging approach, similar to the one provided in Section \ref{sec:FrameworkCalibrationVarMin}. The hedge ratio is calibrated with realisations of $C^{(R)}$ and $C^{(H)}$, denoted as $h^{(R)}$ for the reference population $R$. Therefore, the hedged position contains $C^{(H)}$ and $h^{(R)}C^{(R)}$. The graphical risk metric constructs multiple joint prediction regions, reflecting varying uncertainty levels, defined by deviations between $C^{(H)}$ and $h^{(R)}C^{(R)}$. Formally, the joint prediction region $\mathbf{J}_{\alpha}$ for $(C^{(H)}, h^{(R)}C^{(R)})$ is defined as
    \begin{equation*}
        \Pr \left\{ \left(C^{(H)}, h^{(R)}C^{(R)} \right) \in \mathbf{J}_{\alpha} \right\} = 1 - \alpha,
    \end{equation*}
    where $\alpha \in [0,1]$ is the level of uncertainty, such that the area of $\mathbf{J}_{\alpha}$ contains $100(1-\alpha)$\% of the realisations of $(C^{(H)}, h^{(R)}C^{(R)})$. A higher level of population basis risk corresponds to a larger area spanned by $\mathbf{J}_{\alpha}$ for a given $\alpha$.
    
    Finally, the graphical population basis risk metric is constructed as follows:
    \begin{enumerate}
        \item Simulate $N$ realisations of the mortality rates that are relevant to $S^{(H)}$ and $S^{(R)}$, and use the simulated mortality rates to calculate $N$ realized values of $C^{(H)} = S^{(H)} - \mathbb{E}(S^{(H)})$ and $C^{(R)} = S^{(R)} - \mathbb{E}(S^{(R)})$.
        \item Use the realized values of $C^{(H)}$ and $C^{(R)}$ to compute the hedge ratio $h^{(R)}$, and obtain the hedged position consisting of $C^{(H)}$ and $h^{(R)}C^{(R)}$.
        \item For each realisation of $\mathbf{Y} := \left(C^{(H)}, h^{(R)}C^{(R)}\right)^{'}$, calculate the Mahalanobis distance to the origin as $\mathbf{Y}^{'}\mathbf{\Sigma}^{-1}\mathbf{Y}$, where $\mathbf{\Sigma}$ is the covariance matrix of $\mathbf{Y}$ and is estimated using the $N$ realisations of $\mathbf{Y}$.
        \item Order the $N$ realisations of $\mathbf{Y}$ by their Mahalanobis distance, and select the observations that have the shortest $N(1-\alpha)$ distances with $\alpha = 0.1, 0.2, \dots, 0.9$.
        \item Enclose the selected $N(1-\alpha)$ observations of $\mathbf{Y}$ by drawing a convex hull to form the joint prediction region for each $\alpha$, where different values of $\alpha$ will be given different levels of shading transparency.
    \end{enumerate}

\subsection{Construction of the graphical risk metric} \label{append:grm_construction}

    In this appendix, we provide the procedure for constructing the proposed graphical risk metric for natural hedging. Recall from Table \ref{tab:Valuation} that $\mathcal{A}$ and $\mathcal{I}$ are the present value random variables of the annuity and insurance portfolios, respectively, in the valuation step. The present value of the calibrated insurance portfolio is defined as $\mathcal{L} := \mathcal{L}(h) = h\cdot\mathcal{I}$, where $h$ is the hedge ratio computed in the calibration step. At the evaluation step, the hedged portfolio is expressed as $\mcP(h) = \mathcal{A} + \mathcal{L}(h)$. 
    
    \begin{figure}[!ht]
        \centering
        \begin{subfigure}[t]{0.475\textwidth}
            \centering
            \includegraphics[width=\linewidth]{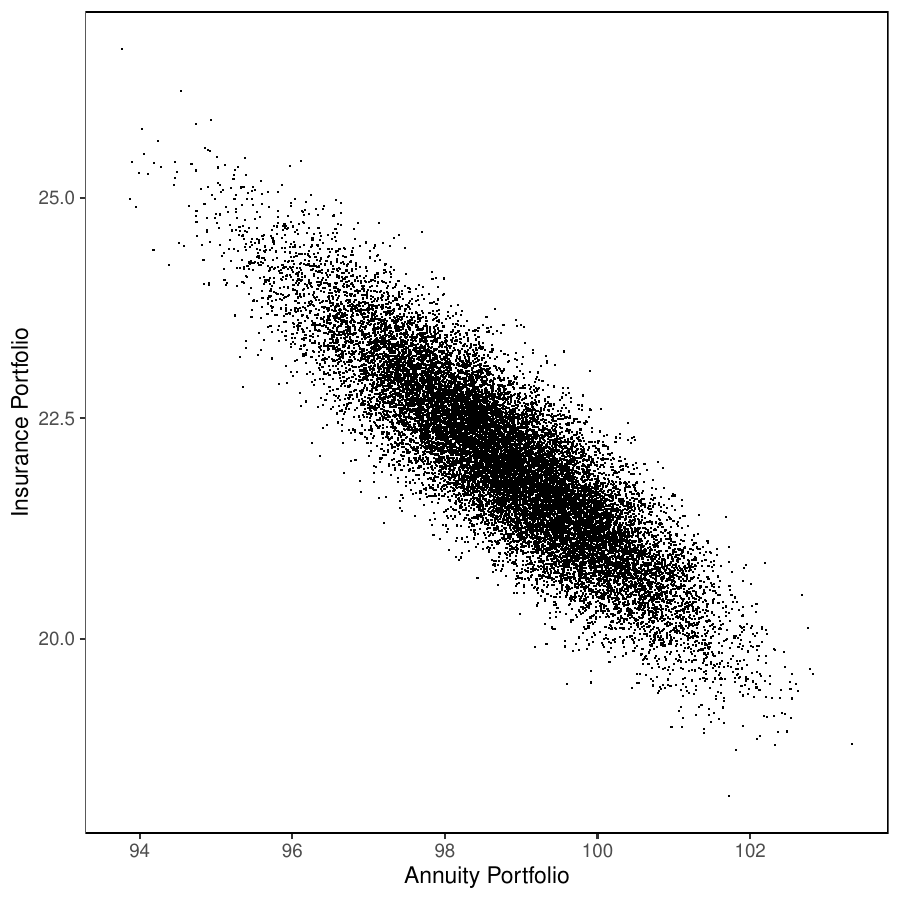}

            \caption{\footnotesize A scatter plot of 20,000 realisations of the assumed annuity and insurance portfolios, $(\mathcal{A},\mathcal{L})$.}
            \label{fig:Method_1}
        \end{subfigure}
        \hfill
        \begin{subfigure}[t]{0.475\textwidth}
            \centering
            \includegraphics[width=\linewidth]{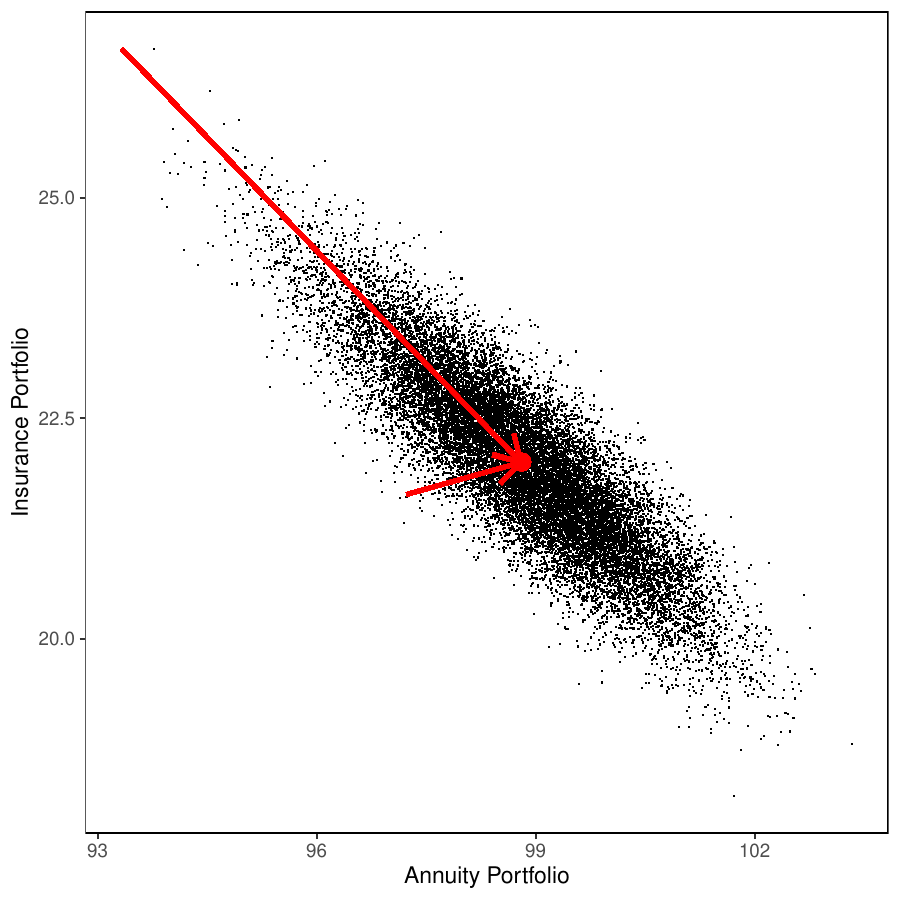}
            \caption{\footnotesize Two red arrow lines indicating the physical distance from a realisation of $(\mathcal{A},\mathcal{L})$ to the expectation point $(A,L)$.}
            \label{fig:Method_2}
        \end{subfigure}
        \vspace{0.25cm}
        \begin{subfigure}[t]{0.475\textwidth}
            \centering
            \includegraphics[width=\linewidth]{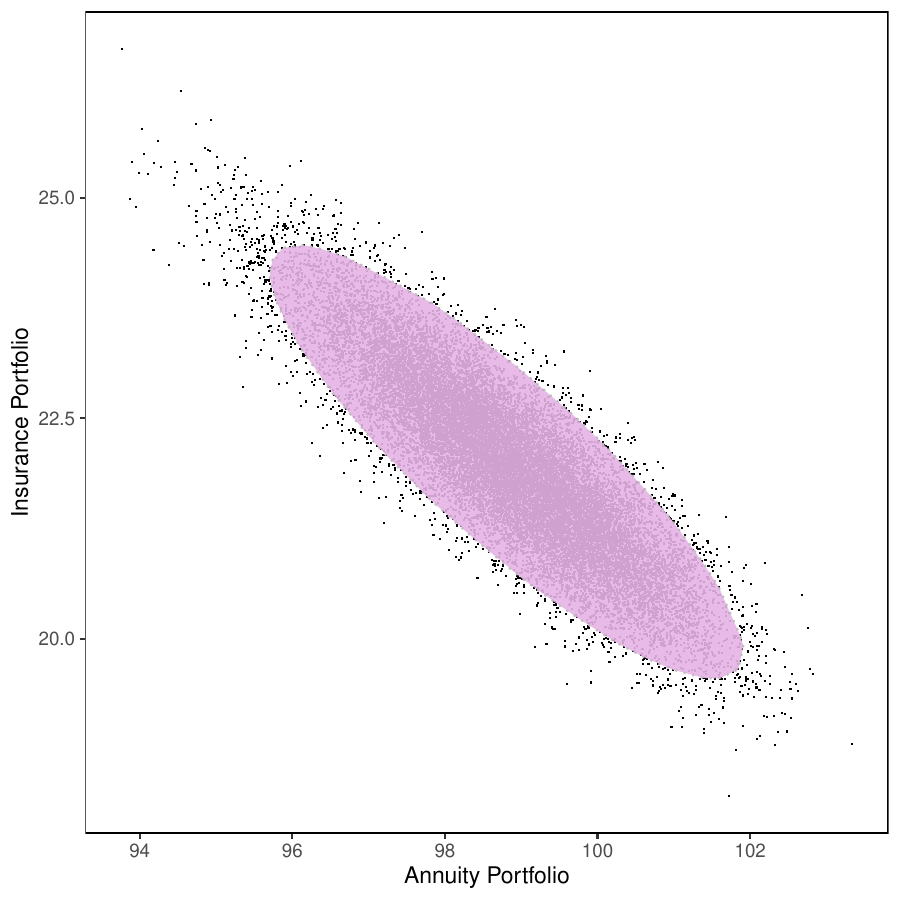}
            \caption{\footnotesize Convex hull for $\mathbf{J}_{0.05}$, the joint prediction region that contains 95\% of the 20,000 realisations of $(\mathcal{A},\mathcal{L})$.}
            \label{fig:Method_3}
        \end{subfigure}
        \hfill
        \begin{subfigure}[t]{0.475\textwidth}
            \centering
            \includegraphics[width=\linewidth]{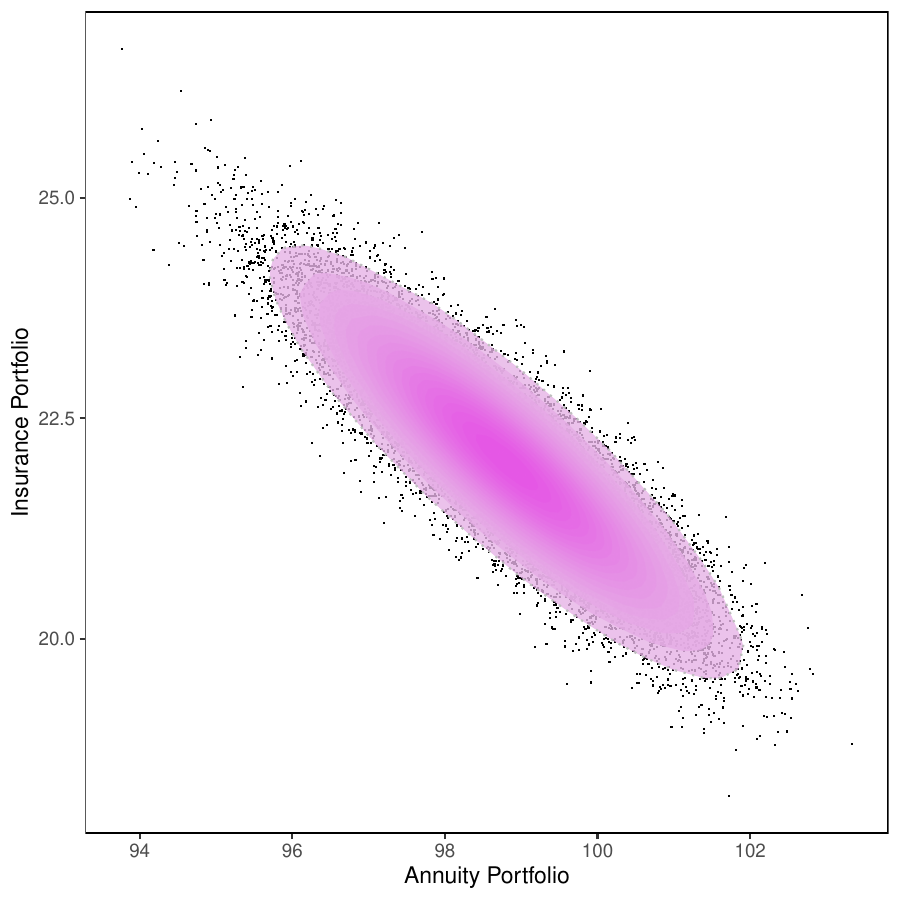}
            \caption{\footnotesize A series of convex hulls for $\mathbf{J}_{\alpha}$, $\alpha = 0.05, 0.10, \dots, 0.95$, with a darker degree of shading as $\alpha$ increases.}
            \label{fig:Method_4}
        \end{subfigure}
        \caption{The construction of graphical risk metric for natural hedging.}
        \label{fig: grm_method_one_port}
    \end{figure}

    An objective of our graphical risk metric is to visualize the joint distribution of $(\mathcal{A},\mathcal{L})$. Figure \ref{fig: grm_method_one_port} presents the step-by-step progression of a graphical region construction for a single portfolio. This portfolio is from the toy example with an issuing age of 40 for life insurance. Figure \ref{fig:Method_1} shows a scatter plot of 20,000 realisations of $(\mathcal{A},\mathcal{L})$, generated by the Lee-Carter model. We observe an inverse relationship between $\mathcal{A}$ and $\mathcal{L}$ -- as $\mathcal{A}$ increases, $\mathcal{L}$ generally decreases, forming a downward sloping cloud of points. This relationship is expected, as both variables depend on survival probabilities in opposite ways; when projected survival probabilities rise, the present value of future annuity liabilities increase, while the present value of future insurance liabilities decrease. 
   
    Following \cite{graphical_basis_risk_measure}, we measure the Mahalanobis distance for each pair $(\mathcal{A},\mathcal{L})$ to its expectation value $(A,L)$, where $A=\bbE[\mathcal{A}]$ and $L=\bbE[\mathcal{L}]$. The Mahalanobis distance of the realisations $\bm{\mathcal{Y}} := (\mathcal{A},\mathcal{L})'$ to $\mathbf{Y} := (A,L)'$ is calculated as
    \[
        \left(\bm{\mathcal{Y}} - \bm{Y}\right)^{'}\bm{\Sigma}^{-1}\left(\bm{\mathcal{Y}} - \bm{Y} \right), 
    \]
    where $\bm{\Sigma}$ is the covariance matrix of $\bm{\mathcal{Y}}$. In \autoref{fig:Method_2}, the red dot at the center of the cloud represents the position of $\mathbf{Y}$, while the two red lines depicts the physical distance of the realisations of $\bm{\mathcal{Y}}$ to $\mathbf{Y}$. The Mahalanobis distance can be interpreted as the physical distance weighted by the covariance matrix $\bm{\Sigma}$.
   
    We are now ready to construct the graphical risk metric. Similar to \cite{graphical_basis_risk_measure}, our proposed graphical risk metric consists of multiple joint prediction regions representing different levels of uncertainty. For a probability $\alpha \in [0,1]$, the joint prediction region $\mathbf{J}_{\alpha}$ for $(\mathcal{A},\mathcal{L})$ is given by
    \begin{equation*}
        \Pr \left\{ (\mathcal{A},\mathcal{L}) \in \mathbf{J}_{\alpha} \right\} = 1 - \alpha.
    \end{equation*}
    \autoref{fig:Method_3} displays the joint prediction region for $\alpha = 0.05$, where $\mathbf{J}_{0.05}$ contains 95\% of the 20,000 realisations of $(\mathcal{A},\mathcal{L})$.

    \autoref{fig:Method_4} displays multiple joint prediction regions where different levels of shading distinguish the varying values of $\alpha$. As $\alpha$ increases, the corresponding joint prediction region is shaded darker. \autoref{fig:Method_4} illustrates the gradient effect of joint prediction regions for $\alpha = 0.05, 0.10, \dots, 0.95$. The layers of joint prediction regions form an oval-shape object, with the smallest region at the center containing 5\% of the realisations and the largest region, with the lightest shading, containing 95\% of the realisations.
    
    We outline the procedure for constructing a series of joint prediction regions for the graphical risk metric for natural hedging: 
    \begin{enumerate}
        \item Simulate $N$ realisations of $\mathcal{A}$ and $\mathcal{L}$ using a selected mortality scenario generator, such as the Lee-Carter model.
        \item Calculate the Mahalanobis distance for each realisation of $(\mathcal{A},\mathcal{L})$ to its empirical expectation.
        \item Sort the $N$ realisations of $(\mathcal{A},\mathcal{L})$ by their Mahalanobis distance, and subset the observations with the shortest $N(1-\alpha)$ distances for $\alpha = 0.05, 0.1, \dots, 0.95$.
        \item Draw a convex hull that encloses the selected $N(1-\alpha)$ observations for each $\alpha$, adjusting the shading level as $\alpha$ changes.
    \end{enumerate}

\section{Mortality Scenario Generator} \label{append:mort_generator}

\subsection{Model-based approach} \label{append:mort_model_sim}
    To simulate survival probabilities under a stochastic mortality model, we apply the following procedure:
    \begin{enumerate}
        \item \textbf{Data}: Import data into \texttt{R} from the Human Mortality Database using the \textit{HMDHFDplus} \texttt{R} package \citep{HMDHFDplus_Rpackage}. We imported central death rates for lives aged $x$ in year $t$, denoted as $m_{x, t}$, from the US male population with years 1970-2018 and age 40-99.
            
        \item \textbf{Fitting}: Use the \textit{StMoMo} package in \texttt{R} \citep{stmomo} to:
        \begin{enumerate}
            \item Define an StMoMo object for the selected model. For the Lee-Carter model, define the object by \texttt{lc(link = ``log'', const = ``sum'')}. For the Cairns-Blake-Dowd model, define the object by \texttt{cbd(link = ``logit'')}.
            \item Fit the model using the \texttt{fit()} function, where the argument includes the StMoMo model object and the mortality data to be fitted.
        \end{enumerate}
        \item \textbf{Projection}: Project mortality rates using the \texttt{simulate()} function, where the argument includes the fitted model, the number of sample paths, and the projection horizon. The output consists of projected central death rates $m_{x,t}$ for the Lee-Carter model and probabilities of death $q_{x,t}$ for the Cairns-Blake-Dowd model
        \item \textbf{Calculation}:
        Substitute the projected $m{x,t}$ values into equation \eqref{eq: gen_survival_prob} to obtain simulated survival probabilities $\mathcal{S}_{x}(T)$. For the Cairns-Blake-Dowd model, we convert $q_{x,t}$ into $m_{x, t}$ using the constant force of mortality assumption.
    \end{enumerate}
    
\subsection{Model-free approach}\label{append:bootstrap_sim}
    To generate stochastic survival probabilities from a non-parametric bootstrapping method, we follow the procedure outlined in \cite{bootstrap}.
    \begin{enumerate}
        \item \textbf{Data}: Import data into \texttt{R} from the Human Mortality Database using the \textit{HMDHFDplus} \texttt{R} package \citep{HMDHFDplus_Rpackage}. We imported central death rates for lives aged $x$ in year $t$, denoted as $m_{x, t}$, from the US male population with years 1970-2018 and age 40-99.
        \item \textbf{Reduction calculation}:
        \begin{enumerate}
            \item Obtain historical mortality reduction rates for lives aged $x$ in year $t$, denoted as $r_{x,t} = \frac{m_{x,t+1}}{m_{x,t}}$.
            Since the dataset spans 49 years (1970-2018), we have a total of 48 reduction rates for each age.
            \item Construct a mortality reduction matrix, where each row corresponds an age $x$ and each column denotes a year $t$. The mortality reduction matrix from our data is
            \begin{equation*}
                \mathbf{r} = \begin{bmatrix}
                    r_{40, 1970} & r_{40, 1971} & \dots & r_{40, 2017} \\
                    r_{41, 1970} & r_{41, 1971} & \dots & r_{41, 2017} \\
                    \vdots & \vdots & \ddots & \vdots \\
                    r_{99, 1970} & r_{99, 1971} & \dots & r_{99, 2017}
                \end{bmatrix}.
            \end{equation*}
            \item Form overlapping mortality reduction blocks of size two, following \cite{bootstrap}. We obtained 47 mortality reduction blocks, corresponding to $t = 1970, \dots, 2017$: $
            \left(\mathbf{r}_{1970}, \mathbf{r}_{1971} \right), \left(\mathbf{r}_{1971}, \mathbf{r}_{1972} \right), \dots, \left(\mathbf{r}_{2016}, \mathbf{r}_{2017} \right)$.
        \end{enumerate}
        \item \textbf{Pseudo sampling}:
        \begin{enumerate}
            \item Set the initial mortality rates to those from the most recent year in the dataset. We set the initial rates to $m_{x, 2018}$ for $x=40,\dots,99$.
            \item Sample with replacement from the mortality reduction blocks to form a pseudo reduction matrix of future mortality. We sampled 35 blocks with replacement to project mortality for 70 future years.
            \item Calculate the projected mortality rates by multiplying the initial rates across the pseudo reduction matrix. We calculated the projected values of $m_{x,t}$ for $x=40,\dots,99$ and $t=2019, \dots, 2089$.
            \item Repeat Steps 3(a)-3(c) to obtain multiple sets of projected mortality rates. We repeated these steps 20,000 times.
        \end{enumerate}
        \item \textbf{Calculation}:
        Substitute the projected $m_{x,t}$ values into equation \eqref{eq: gen_survival_prob} to compute simulated survival probabilities $\mathcal{S}_{x}(T)$.
    \end{enumerate}


\section{Numerical Illustration Details} \label{append:ill_weights}

    This appendix summarises the policy specifications used in the numerical illustrations presented in Section~\ref{sec:illustrations}. Table~\ref{tab:ill_details} outlines the issuing ages $x$, terms $t$, deferrals $\tau$ and weights $\omega$ of the annuity and insurance portfolios considered in the three illustrations. Specifically, Annuity~Portfolio~1 and Insurance~Portfolios~1-3 are used in Illustration~1, Annuity~Portfolio~2 along with Insurance~Portfolio~1 are used in Illustration~2, and Annuity~Portfolio~1 together with Insurance~Portfolio~1 are used in Illustration~3.

    \begin{table}[!ht]
    \centering
    \setlength{\tabcolsep}{3pt}
    \begin{tabular}{c| ccc c|ccc c|ccc|ccc|ccc}
    \toprule
    \multirow{2}{*}{$j$}
     & \multicolumn{4}{c|}{\textbf{Annuity 1}} 
     & \multicolumn{4}{c|}{\textbf{Annuity 2}} 
     & \multicolumn{3}{c|}{\textbf{Insurance 1}} 
     & \multicolumn{3}{c|}{\textbf{Insurance 2}} 
     & \multicolumn{3}{c}{\textbf{Insurance 3}} \\ 
    \cmidrule(lr){2-5} \cmidrule(lr){6-9} \cmidrule(lr){10-12} \cmidrule(lr){13-15} \cmidrule(lr){16-18}
     & $x_j$ & $\omega_j$ & $\tau_j$ & $t_j$
     & $x_j$ & $\omega_j$ & $\tau_j$ & $t_j$
     & $x_j$ & $\omega_j$ & $t_j$
     & $x_j$ & $\omega_j$ & $t_j$
     & $x_j$ & $\omega_j$ & $t_j$ \\ 
    \midrule
     1  & 40 & 0.0517 & 25 & 35 & 40 & 0.0517 & 25 & 20 & 40 & 0.0517 & 60 & 40 & 0.1084 & 60 & 40 & 0.1084 & 20 \\
     2  & 41 & 0.0512 & 24 & 35 & 41 & 0.0512 & 24 & 20 & 41 & 0.0512 & 59 & 41 & 0.1073 & 59 & 41 & 0.1073 & 20 \\
     3  & 42 & 0.0506 & 23 & 35 & 42 & 0.0506 & 23 & 20 & 42 & 0.0506 & 58 & 42 & 0.1062 & 58 & 42 & 0.1062 & 20 \\
     4  & 43 & 0.0494 & 22 & 35 & 43 & 0.0494 & 22 & 20 & 43 & 0.0494 & 57 & 43 & 0.1035 & 57 & 43 & 0.1035 & 20 \\
     5  & 44 & 0.0477 & 21 & 35 & 44 & 0.0477 & 21 & 20 & 44 & 0.0477 & 56 & 44 & 0.1000 & 56 & 44 & 0.1000 & 20 \\
     6  & 45 & 0.0467 & 20 & 35 & 45 & 0.0467 & 20 & 20 & 45 & 0.0467 & 55 & 45 & 0.0980 & 55 & 45 & 0.0980 & 20 \\
     7  & 46 & 0.0456 & 19 & 35 & 46 & 0.0456 & 19 & 20 & 46 & 0.0456 & 54 & 46 & 0.0955 & 54 & 46 & 0.0955 & 20 \\
     8  & 47 & 0.0450 & 18 & 35 & 47 & 0.0450 & 18 & 20 & 47 & 0.0450 & 53 & 47 & 0.0942 & 53 & 47 & 0.0942 & 20 \\
     9  & 48 & 0.0447 & 17 & 35 & 48 & 0.0447 & 17 & 20 & 48 & 0.0447 & 52 & 48 & 0.0937 & 52 & 48 & 0.0937 & 20 \\
    10  & 49 & 0.0445 & 16 & 35 & 49 & 0.0445 & 16 & 20 & 49 & 0.0445 & 51 & 49 & 0.0932 & 51 & 49 & 0.0932 & 20 \\
    11  & 50 & 0.0458 & 15 & 35 & 50 & 0.0458 & 15 & 20 & 50 & 0.0458 & 50 &  –  & – & – & – & – & – \\
    12  & 51 & 0.0484 & 14 & 35 & 51 & 0.0484 & 14 & 20 & 51 & 0.0484 & 49 &  –  & – & – & – & – & – \\
    13  & 52 & 0.0494 & 13 & 35 & 52 & 0.0494 & 13 & 20 & 52 & 0.0494 & 48 &  –  & – & – & – & – & – \\
    14  & 53 & 0.0480 & 12 & 35 & 53 & 0.0480 & 12 & 20 & 53 & 0.0480 & 47 &  –  & – & – & – & – & – \\
    15  & 54 & 0.0464 & 11 & 35 & 54 & 0.0464 & 11 & 20 & 54 & 0.0464 & 46 &  –  & – & – & – & – & – \\
    16  & 55 & 0.0458 & 10 & 35 & 55 & 0.0458 & 10 & 20 & 55 & 0.0458 & 45 &  –  & – & – & – & – & – \\
    17  & 56 & 0.0461 & 9  & 35 & 56 & 0.0461 & 9  & 20 & 56 & 0.0461 & 44 &  –  & – & – & – & – & – \\
    18  & 57 & 0.0471 & 8  & 35 & 57 & 0.0471 & 8  & 20 & 57 & 0.0471 & 43 &  –  & – & – & – & – & – \\
    19  & 58 & 0.0484 & 7  & 35 & 58 & 0.0484 & 7  & 20 & 58 & 0.0484 & 42 &  –  & – & – & – & – & – \\
    20  & 59 & 0.0488 & 6  & 35 & 59 & 0.0488 & 6  & 20 & 59 & 0.0488 & 41 &  –  & – & – & – & – & – \\
    21  & 60 & 0.0487 & 5  & 35 & 60 & 0.0487 & 5  & 20 & 60 & 0.0487 & 40 &  –  & – & – & – & – & – \\
    \bottomrule
    \end{tabular}
    \caption{Specifications and weights for the annuity and insurance portfolios used in the numerical illustrations.}
    \label{tab:ill_details}
    \end{table}

    All portfolio weights are constructed using population counts from the U.S. male population in year  2023, obtained from the Human Mortality Database via the \textit{HMDHFDplus} package \citep{HMDHFDplus_Rpackage}. For each issuing age $x$, the weight $\omega_x$ is calculated as the proportion of the population at age $x$ relative to the total population across all ages considered in that portfolio:
    \[
    \omega_x = \frac{\text{Population count at age } x}{\sum_{x' \in \mathcal{X}} \text{Population count at age } x'},
    \]
    where $\mathcal{X}$ is the set of all ages considered.

\end{document}